\documentclass{article}
\usepackage{amsmath}
\usepackage{a4wide}

\newcommand{\alvp}{a_s}
\newcommand{\fl}{\mathrm{fl}}
\newcommand{\Break}{ \nonumber \\ & & }

\catcode`\@=11
\def\slash{\mathpalette\make@slash}
\def\make@slash#1#2{\setbox\z@\hbox{$#1#2$}%
  \hbox to 0pt{\hss$#1/$\hss\kern-\wd0}\box0}
\catcode`\@=12 %

\begin{document}
\thispagestyle{empty}
\title{
   \vskip-3cm{\baselineskip14pt
   \centerline{\normalsize\hfill TTP00-13}
   \centerline{\normalsize\hfill NIKHEF-2000-018}
  }
 \vskip.7cm
 {\Large\bf{Some higher moments of deep inelastic structure functions
 at next-to-next-to leading order of perturbative QCD}
 \vspace{1.5cm}
 }
}

\author{ A. R\'etey${}^a$ and J.A.M. Vermaseren${}^b$
  \\[2em]
  ${}^a$  {\it Institut f\"ur Theoretische Teilchenphysik,} \\
  {\it Universit\"at Karlsruhe, D-76128 Karlsruhe, Germany} \\
  ${}^b$  {\it NIKHEF Theory Group }\\
        {\it Kruislaan 409, 1098 SJ Amsterdam, The Netherlands}
}

\date{}
\maketitle

\begin{abstract}
\noindent We present the analytic next-to-next-to-leading QCD
calculation of some higher moments of deep inelastic structure
functions in the leading twist approximation. We give results for the
moments $N$=1,3,5,7,9,11,13 of the structure function $F_3$. Similarly we 
present the moments $N$=10,12 for the flavour singlet and $N$=12,14 for the 
non-singlet structure functions $F_2$ and $F_L$. We have calculated both 
the three-loop anomalous dimensions of the corresponding operators and the 
three-loop coefficient functions of the moments of these structure 
functions.
\end{abstract}

\setcounter{page}{1}
\newpage 

\section{Introduction}

The determination of the next-to-next-to-leading (NNL) order QCD
approximation for the structure functions of deep inelastic scattering
has become important for the understanding of perturbative QCD and necessary
for an accurate comparison of perturbative QCD with the increasing
precision of experiments. Such calculations however are rather 
complicated and hence a complete NNL result does not exist as of yet. 
The one-loop anomalous dimensions were calculated in
\cite{phrva:d8:3633,phrva:d9:980}. In \cite{phrva:d18:3998} (see also
the references therein) the complete one-loop coefficient functions
were obtained. Anomalous dimensions at 2-loop order were obtained in
\cite{nupha:b129:66,nupha:b139:545,nupha:b152:493,nupha:b153:161,
nupha:b166:429,nupha:b183:157,phlta:b97:437,nupha:b175:27,nupha:b379:143}. 
The 2-loop coefficient functions were calculated in
\cite{phrva:d25:71,phrva:d30:541,nupha:b307:721,phlta:b291:171,
prlta:65:1535,phlta:b272:127,phlta:b273:476,nupha:b383:525},
\cite{hep-ph/9912355}.
 
Analytical results of the 3-loop anomalous dimensions and coefficient
functions of the moments of $F_2$ and $F_L$ are only known for the
moments $N$=2,4,6,8 in both the singlet and non-singlet case and
additionally for $N$=10 in the non-singlet case from
\cite{nupha:b427:41} and \cite{nupha:b492:338}. In addition the 
Gross-Llevellyn Smith sum rule, which corresponds to the first moment of 
$F_3^{\nu\mathrm{p}+\bar\nu\mathrm{p}}$ has been calculated at this 
order~\cite{phlta:b259:345}.

For a complete reconstruction of the $x$-dependence of the structure 
functions via an inverse Mellin-transformation one would need the moments 
for all $N$ (that is either all even or all odd integer values). 
Additionally one needs them both for $F_2$ and $F_3$ in order to untangle 
the various quark and gluon contributions. The determination of the NNL 
approximation for generic $N$ is work in progress~\cite{hep-ph/0004235}, 
but probably will not be finished in the near future. 

The available moments of $F_2$ have been used by a number of authors to 
make a reconstruction of the complete structure functions at NNL by a 
variety of means~\cite{nupha:b492:338,phlta:b333:190,phlta:b388:179,
hep-ph/9907472}. Additionally they can be used to obtain a better value of 
$\alpha_S$~\cite{nupha:b563:45}
It should be clear that it is important to have as large a number of 
moments as possible. First, these results
can be immediately used to increase the precision of phenomenological
investigations of deep inelastic scattering
\cite{hep-ph/0006154}. Second, the moments will be a very important
check for the new methods and programs needed for the determination of
the 3-loop results for arbitrary $N$.
Unfortunately it is not 
very easy to increase the number of moments, because each new moment 
requires roughly five times the computer resources that its 
predecessor needs. With the advent of better computers this means that by now 
it has been possible to obtain two more moments for the singlet case and 
one additional moment for the non singlet $F_2$ case. This should allow 
for instance a somewhat better determination of $\alpha_S$. More 
important however is the determination of the first seven odd moments of 
$F_3^{\nu\mathrm{p}+\bar\nu\mathrm{p}}$ to three loops. 
To this end we used the the same programs as in
\cite{nupha:b492:338}, state of the art computers and a new version of
the symbolic manipulation program FORM \cite{prog:form} that supports
now 64-bit architectures and to some extend parallel computers (see
also \cite{hep-ph/9906426}). We could push the limit in these
calculations to include two new moments ($N$=10,12) in the calculation
of the flavour singlet structure functions $F_L$ and
$F_2$, and two new moments ($N$ = 12,14) for the flavour nonsinglet structure 
functions $F_L$ and $F_2$. Additionally we have computed the moments 
$N$=3,5,7,9,11,13 of the structure function $F_3$. We do not expect more 
moments to become available before the complete results for all $N$ will 
be presented.

\section{The formalism}

This calculation follows the one presented in \cite{nupha:b492:338}
(see also \cite{hep-ph/9912355}) in every detail, so we only will give
a very short review on the methods used.

We need to calculate the hadronic part of the amplitude for
unpolarized deep inelastic scattering which is given by the hadronic
tensor 
\begin{eqnarray}
\label{def:hadrontensor}
W_{\mu\nu}(x,Q^2) & = & 
\frac{1}{4\pi} \int \mathrm{d}^4 z \mathrm{e}^{\mathrm{i}q\cdot z} 
\langle p,\mathrm{nucl} | J_\mu(z)J_\nu(0)| \mathrm{nucl},p \rangle 
\nonumber \\ & = & 
(g_{\mu\nu} - q_\mu q_\nu )\frac{1}{2x} F_L(x,Q^2) 
\nonumber \\ & &
+ \left( - g_{\mu\nu} -p_\mu p_\nu \frac{4x^2}{q^2} 
- (p_\mu q_\nu + p_\nu q_\mu)\frac{2x}{q^2} \right) \frac{1}{2x} F_2(x,Q^2) 
\nonumber \\ & &
+ \mathrm{i} \epsilon_{\mu\nu\rho\sigma} \frac{p^\rho
q^\sigma}{p\cdot q} F_3(x,Q^2)
\end{eqnarray}
where the $J^\mu$ are either electromagnetic or weak hadronic currents
and $x=Q^2/(2\,p\cdot q)$ is the Bjorken scaling variable with $0 < x
\leq 1$. $Q^2=-q^2$ is the transfered momentum and
$|p,\mathrm{nucl}\rangle$ is the nucleon state with momentum $p$. In
these equations spin averaging is assumed. The longitudinal structure
function $F_L$ is related to the structure function $F_1$ by $F_L =
F_2 - 2 x F_1$. For electron-nucleon scattering $J^\mu$ is the
electromagnetic quark current and $F_3$ vanishes. For neutrino-nucleon
scattering $J^\mu$ is an electroweak quark current which has an axial
vector contribution and $F_3$, which describes parity violating
effects that arise from vector and axial-vector interference, will not
vanish.

Using the dispersion relation technique one can relate the hadronic
tensor to the following 4-point Green functions:
\begin{equation*}
W^{\mu\nu}(p,q) = \frac{1}{2\pi} \mathrm{Im} T^{\mu\nu}(p,q),\quad
T_{\mu\nu}(p,q) = \mathrm{i} \int \mathrm{d}^4 z 
\mathrm{e}^{\mathrm{i}q\cdot z} 
\langle p,\mathrm{nucl} | T\left[ J_\mu(z)J_\nu(0)\right]| 
\mathrm{nucl},p \rangle .
\end{equation*}
Applying a formal operator product expansion in terms of local
operators to the time-ordered product of two quark currents leads to:
\begin{equation}
\label{def:ope}
\begin{split}
\mathrm{i} \int \mathrm{d}^4z \mathrm{e}^{\mathrm{i}q\cdot z}
T \left[ J_{\nu_1}(z) J_{\nu_2}(0) \right] =
\sum_{N,j}\left(\frac{1}{Q^2}\right)^N
\left[\left(g_{\nu_1\nu_2}-\frac{q_{\nu_1}q_{\nu_2}}{q^2}\right)
q_{\mu_1}q_{\mu_2} C^{j}_{L,N}(\frac{Q^2}{\mu^2},a_s)\right.\\
-\left(g_{\nu_1\mu_1}g_{\mu_2\nu_2}q^2
-g_{\nu_1\mu_1}q_{\nu_2}q_{\mu_2}
-g_{\nu_2\mu_2}q_{\nu_1}q_{\mu_1}
+g_{\nu_1\nu_2}q_{\mu_1}q_{\mu_2}\right)
C^{j}_{2,N}(\frac{Q^2}{\mu^2},a_s)\\\left.
+ \mathrm{i}\, \epsilon_{\nu_1\nu_2\mu_1\nu_3} g^{\nu_3\nu_4}
q_{\nu_4} q_{\mu_2} C^{j}_{3,N}(\frac{Q^2}{\mu^2},a_s)
\right] \times 
q_{\mu_3} \dots q_{\mu_N} O^{j,\{\mu_1,\dots\mu_N\}}(0)
+ \mathrm{higher} \: \mathrm{twists},\\ \qquad j = \alpha,\psi,G
\end{split}
\end{equation}
Here we have introduced the notation 
$a_s=\alpha_s/(4\pi)=g^2/(4\pi)^2$ and everything is assumed to be
renormalized. The sum over $N$ runs over the standard set of the
spin-$N$ twist-2 irreducible flavour non-singlet quark operators and
the singlet quark and gluon operators:
\begin{align*}
O^{\alpha,\{\mu_1,\dots,\mu_N\}}& = \bar\psi \lambda^\alpha 
\gamma^{\{\mu_1} D^{\mu_2} \cdots D^{\mu_N\}} \psi ,\quad 
\alpha = 1,2,\dots,(n_f^2-1) \\
O^{\Psi,\{\mu_1,\dots,\mu_N\}} & = \bar\psi 
\gamma^{\{\mu_1} D^{\mu_2} \cdots D^{\mu_N\}} \psi \\
O^{G,\{\mu_1,\dots,\mu_N\}} & = G^{\{\mu\mu_1}
D^{\mu_2} \cdots D^{\mu_{N-1}} G^{\mu_N\mu\}}
\end{align*}
Application of this OPE to Eq.~(\ref{def:hadrontensor}) leads to an
expansion for the unphysical values $x\to\infty$. From the proper
analytical continuation to the physical region $0<x\leq 1$ one finds
for the moments of the structure functions $F_2$, $F_L$ and $F_3$:
\begin{align}
\label{moments2ope}
M_{k,N-2} & = \int\limits_{0}^{1} \mathrm{d}x x^{N-2} F_k(x,Q^2) = 
\sum_{i=\alpha,\psi,G} C^i_{k,N}(\frac{Q^2}{\mu^2},a_s) 
A^i_{\mathrm{nucl},N}\,,\quad k=2,L \\
M_{3,N-1} & = \int\limits_{0}^{1} \mathrm{d}x x^{N-1} F_3(x,Q^2) = 
\sum_{i=\alpha} C^{i,N}(\frac{Q^2}{\mu^2},a_s) 
A^{i}_{\mathrm{nucl},N}
\end{align}
with the spin averaged matrix elements
\begin{equation}
\label{def:matrixelements}
\langle p,\mathrm{nucl}| O^{j,\{\mu_1\cdots\mu_N\}} |p,\mathrm{nucl}\rangle
= p^{\{\mu_1}\cdots p^{\mu_N\}} A^j_{\mathrm{nucl},N}(\frac{p^2}{\mu^2}) 
\end{equation}
In the derivation of (\ref{moments2ope}) one needs the symmetry
properties of $T_{\mu\nu}$ under $x\to-x$. This is why one can only
find either even or odd moments from these equations, dependent on the
process under consideration. For $F_3$ we will only consider the
flavor non-singlet contributions, due to the properties of the
operators $O^\psi$ and $O^G$ under charge conjugation there should not
be a singlet contribution (see e.g. \cite{rmpha:52:199}).

The scale-dependence of the coefficient functions is then covered by
the renormalization group equations:
\begin{align}
\label{rge4C}
\left[ 
\mu^2 \frac{\partial}{\partial \mu^2} 
+ \beta(a_s)\frac{\partial}{\partial a_s}
-\gamma_N^{\mathrm{ns}} 
\right] C^{\mathrm{ns}}_{i,N}\left(\frac{Q^2}{\mu^2},a_s\right) & = 0
,\qquad i=2,3,L\\
\sum_{k=\psi,G}\left[ \left\{
\mu^2 \frac{\partial}{\partial \mu^2} 
+ \beta(a_s)\frac{\partial}{\partial a_s}\right\} \delta^{jk}
-\gamma_N^{jk} 
\right] C^{k}_{i,N}\left(\frac{Q^2}{\mu^2},a_s\right) & = 0
,\qquad i=2,L;\, j=\psi,G 
\end{align}
The non-singlet coefficient functions and anomalous dimensions don't
depend on the index $\alpha$, and we have adopted the conventional
collective denotation ``ns'' for them.

\section{The even moments of $F_2$ and $F_L$}

Equation~(\ref{def:ope}) is a relation between operators and does not
depend on the hadronic states to which the OPE is applied.

Using the method of projectors \cite{nupha:b283:452} one can find both
the coefficient functions and the anomalous dimensions for the even
moments of $F_2$ and $F_L$ as defined in Eq.~(\ref{rge4C}) from the
following 4-point Green functions:
\begin{align}
\label{Tqg}
T^{q\gamma q\gamma}_{\mu\nu} & =  \mathrm{i} \int \mathrm{d}^4 z 
\mathrm{e}^{\mathrm{i}q\cdot z} 
\langle p,\mathrm{quark} | T\left[ J_\mu(z)J_\nu(0)\right]| 
\mathrm{quark},p \rangle 
\\
T^{g\gamma g\gamma}_{\mu\nu} & = \mathrm{i} \int \mathrm{d}^4 z 
\mathrm{e}^{\mathrm{i}q\cdot z} 
\langle p,\mathrm{gluon} | T\left[ J_\mu(z)J_\nu(0)\right]| 
\mathrm{gluon},p \rangle
\end{align}
Applying to Eqs.~(\ref{Tqg}) the projectors
\begin{equation}
\label{projector:N}
P_N \equiv \left. \left[ 
\frac{q^{\{\mu_1} \cdots q^{\mu_N\}} }{N!}
\frac{\partial^N}{\partial p^{\mu_1} \cdots \partial p^{\mu_N}}
\right] \right|_{p=0}
\end{equation}
and, to project out the different Lorentz projections (these
projectors are valid in $D=4-2\epsilon$ and for the leading twist
approximation):
\begin{align*}
P_L & = - \frac{q^2}{(p\cdot q)^2} p^\mu p^\nu \\
P_2 & = - \left( 
\frac{3-2\epsilon}{2-2\epsilon} \frac{q^2}{(p\cdot q)^2} p^\mu p^\nu  
+ \frac{1}{2-2\epsilon} g^{\mu\nu} 
\right)\,,
\end{align*}
as well as the corresponding flavour projections results in the
equations:
\begin{align}
\label{singletgreens}
T^{q\gamma q\gamma,\mathrm{s}}_{k,N}(\frac{Q^2}{\mu^2},a_s,\epsilon) & = 
\left( C^{\psi}_{k,N}(\frac{Q^2}{\mu^2},a_s,\epsilon)
Z^{\psi\psi}_N(a_s,\frac{1}{\epsilon}) 
+ C^{G}_{k,N}(\frac{Q^2}{\mu^2},a_s,\epsilon)
Z^{G\psi}_N(a_s,\frac{1}{\epsilon})
\right) & A^{\psi,\mathrm{tree}}_{\mathrm{quark},N}(\epsilon)  
\notag \\
T^{g\gamma g\gamma,\mathrm{s}}_{k,N}(\frac{Q^2}{\mu^2},a_s,\epsilon) & = 
\left( C^{\psi}_{k,N}(\frac{Q^2}{\mu^2},a_s,\epsilon)
Z^{\psi G}_N(a_s,\frac{1}{\epsilon}) 
+ C^{G}_{k,N}(\frac{Q^2}{\mu^2},a_s,\epsilon) 
Z^{GG}_N(a_s,\frac{1}{\epsilon}) 
\right) & A^{G,\mathrm{tree}}_{\mathrm{gluon},N}(\epsilon) 
\notag \\
T^{q\gamma q\gamma,\mathrm{ns}}_{k,N}(\frac{Q^2}{\mu^2},a_s,\epsilon) & = 
C^{\mathrm{ns}}_{k,N}(\frac{Q^2}{\mu^2},a_s,\epsilon)
Z^{\mathrm{ns}}_N(a_s,\frac{1}{\epsilon})
A^{\mathrm{ns},\mathrm{tree}}_{\mathrm{quark},N}(\epsilon)
& k=2,L
\end{align}
From these equations the coefficient functions and from the $Z^{ij}_N$
the anomalous dimensions can be calculated in the usual way.

It should be mentioned that in the Eqs.~(\ref{singletgreens}) on the
left hand side after applying the projectors~(\ref{projector:N}) we
are left with only diagrams of the massless propagator type, a problem
solved at 3-loop order long ago \cite{nupha:b192:159} and implemented
in an efficient way in the FORM package MINCER \cite{prog:mincer}. On
the right hand side only the tree level diagrams contributing to the
Matrix elements survive.

It turns out that much computing time can be saved when calculating
additionally Green functions with external ghosts to get rid of the
unphysical polarization states of the external gluons instead of using
the very complicated projection onto physical states. Also, from
Eqs.~(\ref{singletgreens}) one can determine the $Z^{GG}_N$ and
$Z^{G\psi}_N$ only to order $\alpha_s^2$. To obtain the
$\alpha_s^3$-contributions one can calculate additionally Greens
functions with external scalar fields $\phi$ that couple to gluons
only at tree level. Altogether, to obtain the coefficient functions
and anomalous dimensions for the even moments with $N$=10,12 of $F_2$
and $F_L$ the following diagrams had to be calculated (q=quark,
g=gluon, $\gamma$=photon, h=ghost, $\phi$=scalar field):
\begin{center}
\begin{tabular}{|c|c|c|c|c|c|}
\hline
& tree & 1-loop & 2-loops & 3-loops & Lorentz projections \\
\hline
\hline
$q\gamma q\gamma$ & 1 & 3 & 27 & 413 & 2 \\
$q\phi q\phi$ &  & 1 & 24 & 697 & 1 \\
$g\gamma g\gamma$ &  & 2 & 20 & 366 & 2 \\
$h\gamma h\gamma$ &  &  & 2 & 53 & 2 \\
$g\phi g\phi$ & 1 & 11 & 241 & 1266 & 1 \\
$h\phi h\phi$ &  & 11 & 241 & 1266 & 1 \\
\hline
Total & 3 & 23 & 399 & 10846 & \\
\hline
\end{tabular}
\end{center}
The $q^{\{\mu_1} \cdots q^{\mu_N\}}$ in Eq.~(\ref{projector:N}) are
the harmonic (i.e. symmetrical and traceless) part of the tensor
$q^{\mu_1} \cdots q^{\mu_N}$. The number of terms in these harmonic
tensors explodes as $N$ increases and this is the real limitation in
these calculations considering the computing time as well as
disk-space usage. In spite of a very efficient implementation of these
tensors (see \cite{Larin:1993sc}) for $N=12$, singlet and $N=14$, 
nonsinglet, individual diagrams had a
disk space usage up to and over 100 GB. Altogether the calculation of
all the above diagrams for $N=10,12$ took approximately 5 weeks on a
Compaq Server with 8 Alpha 21264 processors running at 700 MHz, 4 GB
of RAM and 12$\times$17 GB of disk-space. The $N=14$ nonsinglet calculation 
took comparable resources.

\section{The odd moments of $F_3$}

The coefficient functions and anomalous dimensions of the odd moments
of the structure function $F_3$ can be obtained in the same way as the
non-singlet part of $F_2$ and $F_L$ but now considering the
time-ordered product of one vector current $V_\mu$ and one axial
vector current $A_\nu$. The axial current introduces the appearance of
a $\gamma_5$ and some care has to be taken to treat it correctly
within the framework of dimensional regularization. We adopt the
definition used in \cite{phlta:b259:345}:
\begin{equation*}
\gamma_\mu \gamma_5 = \mathrm{i} \frac{1}{6}
\epsilon_{\mu\nu\rho\sigma} \gamma^\mu \gamma^\nu \gamma^\rho
\gamma^\sigma
\end{equation*}
Projecting out the flavour non-singlet part and the corresponding
Lorentz structure with:
\begin{equation*}
P_3 = -\mathrm{i} \frac{1}{(1-2\epsilon)(2-2\epsilon)}
\epsilon^{\mu\nu\alpha\beta} \frac{p_\alpha q_\beta}{p\cdot q}
\end{equation*}
one finds products of metric tensors which have to be considered as
$D$-dimensional objects. Since this definition of $\gamma_5$ in $D$
dimensions violates the axial Ward identity one needs to renormalize
$A_\mu$ with a renormalization constant $Z_A$ and additionally apply a
finite renormalization with $Z_5$, both of these constants are given
to 3-loop order in \cite{phlta:b259:345}.  Combining all this finally
leads to
\begin{equation*}
Z_A(a_s,\frac{1}{\epsilon}) Z_5(a_s,\epsilon)
T^{\mathrm{ns}}_{3,N}(\frac{Q^2}{\mu^2},a_s,\epsilon) =
C_{3,N}(\frac{Q^2}{\mu^2},a_s,\epsilon)
Z_N^{-1}(a_s,\frac{1}{\epsilon})
A^{\mathrm{ns}}_{N,\mathrm{tree}}(\epsilon)
\end{equation*}
Due to the $\gamma_5$ insertion at one of the vertices, some of the
symmetries that were used to minimize the number of diagrams could not
be applied in this case and to determine the $T^{\mathrm{ns}}_{3,N}$
1076 (= 1 + 4 + 55 + 1016) diagrams had to be evaluated, which took about 6
weeks for the moments $N$=1,3,5,7,9,11,13.

\section{Results}

Using the strategies sketched in the previous section we find the
following results for the coefficient functions and anomalous
dimensions. Again following Ref.~\cite{nupha:b492:338}, we present the
combined singlet and non-singlet results for $F_2$ and $F_L$ in terms
of flavour factors which are defined in the following table for $n_f$
number of flavours:
\begin{center}
\begin{tabular}{|l|c|c|c|c|c|}
\hline
\raisebox{-1.5ex}{\rule{0ex}{4ex}}
 & $fl_{2}$ & $fl_{11}$ & $fl_{02}$ & $fl^g_{2}$ & $fl^g_{11}$ \\
\hline
\raisebox{-2.5ex}{\rule{0ex}{5ex}}
non-singlet & 1 & $\frac{3}{n_f} \sum_{f=1}^{n_f} e_f$ & 0 &-&-\\
\raisebox{-2.5ex}{\rule{0ex}{5ex}}
singlet & 1 & 
$\frac{1}{n_f}\frac{\left(\sum_{f=1}^{n_f}e_f\right)^2}{\sum_{f=1}^{n_f}e_f^2}$
& 1 & 1 & 
$\frac{1}{n_f}\frac{\left(\sum_{f=1}^{n_f}e_f\right)^2}{\sum_{f=1}^{n_f}e_f^2}$
\\
\hline
\end{tabular}
\end{center}
The numerical values of the anomalous dimensions for the spin-even
operators contributing to $F_2$ and $F_L$ that now are known to NNL
order are:

\begin{flalign*}
& \nonumber \\
\gamma^{\psi\psi}_{2} & = 
3.555555556\,\alvp + {\alvp^2}\,\left( 48.32921811 - 3.160493827\,n_f - 1.975308642\,\fl_{02}\,n_f \right)  & \nonumber\\ 
 &\qquad+\alvp^3\,\left( 859.4478372 - 133.4381617\,n_f - 1.229080933\,{n_f^2} \right.& \nonumber\\
&\left.\qquad+\fl_{02}\,\left( -42.21182429\,n_f - 3.445816187\,{n_f^2} \right)  \right) 
& \nonumber \\
\gamma^{\psi\psi}_{4} & = 
6.977777778\,\alvp + {\alvp^2}\,\left( 86.28665021 - 6.553580247\,n_f - 0.1060246914\,\fl_{02}\,n_f \right)  & \nonumber\\ 
 &\qquad+\alvp^3\,\left( 1515.562363 - 244.728592\,n_f - 2.108515775\,{n_f^2} \right.& \nonumber\\
&\left.\qquad+\fl_{02}\,\left( -5.17013312\,n_f - 0.6789278464\,{n_f^2} \right)  \right) 
& \nonumber \\
\gamma^{\psi\psi}_{6} & = 
9.003174603\,\alvp + {\alvp^2}\,\left( 108.0184697 - 8.62925674\,n_f - 0.02080408883\,\fl_{02}\,n_f \right)  & \nonumber\\ 
 &\qquad+\alvp^3\,\left( 1891.827779 - 307.4236889\,n_f - 2.570638992\,{n_f^2} \right.& \nonumber\\
&\left.\qquad+\fl_{02}\,\left( -2.526091192\,n_f - 0.2884556035\,{n_f^2} \right)  \right) 
& \nonumber \\
\gamma^{\psi\psi}_{8} & = 
10.45820106\,\alvp + {\alvp^2}\,\left( 123.7764525 - 10.14583662\,n_f - 0.006586485074\,\fl_{02}\,n_f \right)  & \nonumber\\ 
 &\qquad+\alvp^3\,\left( 2164.091836 - 352.3116596\,n_f - 2.882493484\,{n_f^2} \right.& \nonumber\\
&\left.\qquad+\fl_{02}\,\left( -1.682156519\,n_f - 0.1620816452\,{n_f^2} \right)  \right) 
& \nonumber \\
\gamma^{\psi\psi}_{10} & = 
11.5969216\,\alvp + {\alvp^2}\,\left( 136.2741775 - 11.34594534\,n_f - 0.00269944007\,\fl_{02}\,n_f \right)  & \nonumber\\ 
 &\qquad+\alvp^3\,\left( 2379.919952 - 387.6422968\,n_f - 3.115523145\,{n_f^2} \right.& \nonumber\\
&\left.\qquad+\fl_{02}\,\left( -1.256562245\,n_f - 0.1054295071\,{n_f^2} \right)  \right) 
& \nonumber \\
\gamma^{\psi\psi}_{12} & = 
12.53336293\,\alvp + {\alvp^2}\,\left( 146.6771964 - 12.34063169\,n_f - 0.001302012432\,\fl_{02}\,n_f \right)  & \nonumber\\ 
 &\qquad+\alvp^3\,\left( 2559.641948 - 416.9091301\,n_f - 3.300349343\,{n_f^2} \right.& \nonumber\\
&\left.\qquad+\fl_{02}\,\left( -0.9957297081\,n_f - 0.07498848634\,{n_f^2} \right)  \right) 
& \nonumber \\
\gamma^{\psi\psi,NS}_{14} & = 
13.32896733\,\alvp + {\alvp^2}\,\left( 155.6071962 - 13.19066078\,n_f \right)  & \nonumber\\ 
 &\qquad+\alvp^3\,\left( 2714.031720 - 441.9494048\,n_f - 3.452881831\,{n_f^2} \right)
& 
 \end{flalign*}
\begin{flalign*}
\gamma^{\psi G}_{2} & = 
-0.6666666667\,\alvp\,n_f - 7.543209877\,{\alvp^2}\,n_f & \nonumber\\ 
 &\qquad+\alvp^3\,\left( -37.62337275\,n_f + 12.11248285\,{n_f^2} \right) 
& \nonumber \\
\gamma^{\psi G}_{4} & = 
-0.3666666667\,\alvp\,n_f + 1.290703704\,{\alvp^2}\,n_f & \nonumber\\ 
 &\qquad+\alvp^3\,\left( 33.58149273\,n_f + 6.06027262\,{n_f^2} \right) 
& \nonumber \\
\gamma^{\psi G}_{6} & = 
-0.2619047619\,\alvp\,n_f + 2.761104812\,{\alvp^2}\,n_f & \nonumber\\ 
 &\qquad+\alvp^3\,\left( 33.41602135\,n_f + 3.537682102\,{n_f^2} \right) 
& \nonumber \\
\gamma^{\psi G}_{8} & = 
-0.2055555556\,\alvp\,n_f + 3.243957223\,{\alvp^2}\,n_f & \nonumber\\ 
 &\qquad+\alvp^3\,\left( 28.7612615\,n_f + 2.225433112\,{n_f^2} \right) 
& \nonumber \\
\gamma^{\psi G}_{10} & = 
-0.1696969697\,\alvp\,n_f + 3.407168695\,{\alvp^2}\,n_f & \nonumber\\ 
 &\qquad+\alvp^3\,\left( 23.93704198\,n_f + 1.449828678\,{n_f^2} \right) 
& \nonumber \\
\gamma^{\psi G}_{12} & = 
-0.1446886447\,\alvp\,n_f + 3.438705999\,{\alvp^2}\,n_f & \nonumber\\ 
 &\qquad+\alvp^3\,\left( 19.63230379\,n_f + 0.9524545446\,{n_f^2} \right) 
& 
 \end{flalign*}
\begin{flalign*}
\gamma^{G\psi}_{2} & = 
-3.555555556\,\alvp + {\alvp^2}\,\left( -48.32921811 + 5.135802469\,n_f \right)  & \nonumber\\ 
 &\qquad+\alvp^3\,\left( -859.4478372 + 175.649986\,n_f + 4.674897119\,{n_f^2} \right) 
& \nonumber \\
\gamma^{G\psi}_{4} & = 
-0.9777777778\,\alvp + {\alvp^2}\,\left( -16.1752428 + 0.6182716049\,n_f \right)  & \nonumber\\ 
 &\qquad+\alvp^3\,\left( -315.276255 + 39.82571027\,n_f + 1.801843621\,{n_f^2} \right) 
& \nonumber \\
\gamma^{G\psi}_{6} & = 
-0.5587301587\,\alvp + {\alvp^2}\,\left( -9.496317796 + 0.08884857647\,n_f \right)  & \nonumber\\ 
 &\qquad+\alvp^3\,\left( -188.9088124 + 19.67944546\,n_f + 1.087843741\,{n_f^2} \right) 
& \nonumber \\
\gamma^{G\psi}_{8} & = 
-0.3915343915\,\alvp + {\alvp^2}\,\left( -6.757603506 - 0.07061952353\,n_f \right)  & \nonumber\\ 
 &\qquad+\alvp^3\,\left( -134.7055042 + 12.3754454\,n_f + 0.7536013741\,{n_f^2} \right) 
& \nonumber \\
\gamma^{G\psi}_{10} & = 
-0.3016835017\,\alvp + {\alvp^2}\,\left( -5.297576945 - 0.1348941718\,n_f \right)  & \nonumber\\ 
 &\qquad+\alvp^3\,\left( -104.911278 + 8.796702078\,n_f + 0.5579674847\,{n_f^2} \right) 
& \nonumber \\
\gamma^{G\psi}_{12} & = 
-0.2455322455\,\alvp + {\alvp^2}\,\left( -4.398625917 - 0.1639529655\,n_f \right)  & \nonumber\\ 
 &\qquad+\alvp^3\,\left( -86.18107998 + 6.735609285\,n_f + 0.4293379075\,{n_f^2} \right) 
& 
 \end{flalign*}
\begin{flalign*}
\gamma^{GG}_{2} & = 
0.6666666667\,\alvp\,n_f + 7.543209877\,{\alvp^2}\,n_f & \nonumber\\ 
 &\qquad+\alvp^3\,\left( 37.62337275\,n_f - 12.11248285\,{n_f^2} \right) 
& \nonumber \\
\gamma^{GG}_{4} & = 
{\alvp^2}\,\left( 128.178 - 13.64948148\,n_f \right)  + \alvp\,\left( 12.6 + 0.6666666667\,n_f \right)  & \nonumber\\ 
 &\qquad+\alvp^3\,\left( 2066.19278 - 401.3127939\,n_f - 10.43150645\,{n_f^2} \right) 
& \nonumber \\
\gamma^{GG}_{6} & = 
{\alvp^2}\,\left( 183.0538144 - 20.46668466\,n_f \right)  + \alvp\,\left( 17.78571429 + 0.6666666667\,n_f \right)  & \nonumber\\ 
 &\qquad+\alvp^3\,\left( 2987.042058 - 566.6373298\,n_f - 10.78060861\,{n_f^2} \right) 
& \nonumber \\
\gamma^{GG}_{8} & = 
{\alvp^2}\,\left( 219.6240988 - 24.69926432\,n_f \right)  + \alvp\,\left( 21.26666667 + 0.6666666667\,n_f \right)  & \nonumber\\ 
 &\qquad+\alvp^3\,\left( 3609.35419 - 673.9430658\,n_f - 11.20133837\,{n_f^2} \right) 
& \nonumber \\
\gamma^{GG}_{10} & = 
{\alvp^2}\,\left( 247.6655484 - 27.82178573\,n_f \right)  + \alvp\,\left( 23.92337662 + 0.6666666667\,n_f \right)  & \nonumber\\ 
 &\qquad+\alvp^3\,\left( 4089.236943 - 755.1340541\,n_f - 11.57068198\,{n_f^2} \right) 
& \nonumber \\
\gamma^{GG}_{12} & = 
{\alvp^2}\,\left( 270.6428892 - 30.31377688\,n_f \right)  + \alvp\,\left( 26.08168498 + 0.6666666667\,n_f \right)  & \nonumber\\ 
 &\qquad+\alvp^3\,\left( 4483.563048 - 821.1236576\,n_f - 11.88665683\,{n_f^2} \right) 
& 
 \end{flalign*}

The corresponding coefficient functions read:

\begin{flalign*}
C^{\psi}_{2,2} & = 
1 + 0.4444444444\,\alvp + {\alvp^2}\,\left( 17.69376589 - 5.333333333\,n_f - 2.189300412\,\fl_{02}\,n_f \right)  & \nonumber\\ 
 &\qquad+\alvp^3\,\left( 442.7409693 - 165.1971095\,n_f - 24.09201335\,\fl_{11}\,n_f + 6.030272415\,{n_f^2} \right.& \nonumber\\
&\left.\qquad+\fl_{02}\,\left( -79.04486142\,n_f + 3.325504478\,{n_f^2} \right)  \right) 
& \nonumber \\
C^{\psi}_{2,4} & = 
1 + 6.066666667\,\alvp + {\alvp^2}\,\left( 142.3434719 - 16.98791358\,n_f + 0.4858308642\,\fl_{02}\,n_f \right)  & \nonumber\\ 
 &\qquad+\alvp^3\,\left( 4169.267888 - 901.2351626\,n_f - 18.21884618\,\fl_{11}\,n_f + 23.35503924\,{n_f^2} \right.& \nonumber\\
&\left.\qquad+\fl_{02}\,\left( 16.64834849\,n_f - 2.208630689\,{n_f^2} \right)  \right) 
& \nonumber \\
C^{\psi}_{2,6} & = 
1 + 11.17671958\,\alvp + {\alvp^2}\,\left( 302.398735 - 28.0130504\,n_f + 0.4868787285\,\fl_{02}\,n_f \right)  & \nonumber\\ 
 &\qquad+\alvp^3\,\left( 10069.63085 - 1816.322929\,n_f - 16.14271761\,\fl_{11}\,n_f + 42.66273116\,{n_f^2} \right.& \nonumber\\
&\left.\qquad+\fl_{02}\,\left( 24.11778813\,n_f - 1.525489143\,{n_f^2} \right)  \right) 
& \nonumber \\
C^{\psi}_{2,8} & = 
1 + 15.52989418\,\alvp + {\alvp^2}\,\left( 470.807419 - 37.9248228\,n_f + 0.3859585393\,\fl_{02}\,n_f \right)  & \nonumber\\ 
 &\qquad+\alvp^3\,\left( 17162.37245 - 2787.297692\,n_f - 15.09203827\,\fl_{11}\,n_f + 61.91177997\,{n_f^2} \right.& \nonumber\\
&\left.\qquad+\fl_{02}\,\left( 22.33201938\,n_f - 1.036308122\,{n_f^2} \right)  \right) 
& \nonumber \\
C^{\psi}_{2,10} & = 
1 + 19.30061568\,\alvp + {\alvp^2}\,\left( 639.210663 - 46.86131842\,n_f + 0.3045901308\,\fl_{02}\,n_f \right)  & \nonumber\\ 
 &\qquad+\alvp^3\,\left( 24953.13497 - 3770.10212\,n_f - 14.45874451\,\fl_{11}\,n_f + 80.52097973\,{n_f^2} \right.& \nonumber\\
&\left.\qquad+\fl_{02}\,\left( 19.53359559\,n_f - 0.7372434464\,{n_f^2} \right)  \right) 
& \nonumber \\
C^{\psi}_{2,12} & = 
1 + 22.62841097\,\alvp + {\alvp^2}\,\left( 804.5854321 - 54.99446579\,n_f + 0.2451231747\,\fl_{02}\,n_f \right)  & \nonumber\\ 
 &\qquad+\alvp^3\,\left( 33171.45501 - 4746.440949\,n_f - 14.03541028\,\fl_{11}\,n_f + 98.3483124\,{n_f^2} \right.& \nonumber\\
&\left.\qquad+\fl_{02}\,\left( 16.98652635\,n_f - 0.5471547625\,{n_f^2} \right)  \right) 
& \nonumber \\
C^{\psi,NS}_{2,14} & = 
1 + 2.561093284\,\alvp + {\alvp^2}\,\left( 965.8132564 - 62.46549093\,n_f \right)  & \nonumber\\ 
 &\qquad+\alvp^3\,\left( 41657.11568 - 5708.215623\,n_f - 13.73240102\,\fl_{11}\,n_f + 115.3919490\,{n_f^2} \right)
& 
 \end{flalign*}
\begin{flalign*}
C^{G}_{2,2} & = 
-0.5\,\alvp\,n_f - 8.918338961\,{\alvp^2}\,n_f & \nonumber\\ 
 &\qquad+\alvp^3\,\left( -130.7340963\,n_f + 29.37933515\,{n_f^2} - 0.9007972776\,\fl^g_{11}\,{n_f^2} \right) 
& \nonumber \\
C^{G}_{2,4} & = 
-0.7388888889\,\alvp\,n_f - 14.27158692\,{\alvp^2}\,n_f & \nonumber\\ 
 &\qquad+\alvp^3\,\left( -346.4612756\,n_f + 46.52017564\,{n_f^2} - 1.611816512\,\fl^g_{11}\,{n_f^2} \right) 
& \nonumber \\
C^{G}_{2,6} & = 
-0.7051587302\,\alvp\,n_f - 20.06849828\,{\alvp^2}\,n_f & \nonumber\\ 
 &\qquad+\alvp^3\,\left( -715.0372438\,n_f + 61.28545096\,{n_f^2} - 1.496036938\,\fl^g_{11}\,{n_f^2} \right) 
& \nonumber \\
C^{G}_{2,8} & = 
-0.6440873016\,\alvp\,n_f - 23.17873524\,{\alvp^2}\,n_f & \nonumber\\ 
 &\qquad+\alvp^3\,\left( -996.5038709\,n_f + 68.66467304\,{n_f^2} - 1.286400915\,\fl^g_{11}\,{n_f^2} \right) 
& \nonumber \\
C^{G}_{2,10} & = 
-0.5861279461\,\alvp\,n_f - 24.76678064\,{\alvp^2}\,n_f & \nonumber\\ 
 &\qquad+\alvp^3\,\left( -1201.206903\,n_f + 72.23614791\,{n_f^2} - 1.094394334\,\fl^g_{11}\,{n_f^2} \right) 
& \nonumber \\
C^{G}_{2,12} & = 
-0.5358430591\,\alvp\,n_f - 25.51669345\,{\alvp^2}\,n_f & \nonumber\\ 
 &\qquad+\alvp^3\,\left( -1351.047836\,n_f + 73.7936445\,{n_f^2} - 0.9344248731\,\fl^g_{11}\,{n_f^2} \right) 
& 
 \end{flalign*}
\begin{flalign*}
C^{\psi}_{L,2} & = 
1.777777778\,\alvp + {\alvp^2}\,\left( 56.75530152 - 4.543209877\,n_f - 3.950617284\,\fl_{02}\,n_f \right)  & \nonumber\\ 
 &\qquad+\alvp^3\,\left( 2544.598087 - 421.6908885\,n_f - 7.736698288\,\fl_{11}\,n_f + 11.8957476\,{n_f^2} \right.& \nonumber\\
&\left.\qquad+\fl_{02}\,\left( -213.9253076\,n_f + 17.91326528\,{n_f^2} \right)  \right) 
& \nonumber \\
C^{\psi}_{L,4} & = 
1.066666667\,\alvp + {\alvp^2}\,\left( 47.99398931 - 3.413333333\,n_f - 0.6945185185\,\fl_{02}\,n_f \right)  & \nonumber\\ 
 &\qquad+\alvp^3\,\left( 2523.73902 - 383.0520013\,n_f - 5.058869512\,\fl_{11}\,n_f + 10.88895473\,{n_f^2} \right.& \nonumber\\
&\left.\qquad+\fl_{02}\,\left( -55.5530456\,n_f + 2.348005487\,{n_f^2} \right)  \right) 
& \nonumber \\
C^{\psi}_{L,6} & = 
0.7619047619\,\alvp + {\alvp^2}\,\left( 40.9961976 - 2.69569161\,n_f - 0.2524824533\,\fl_{02}\,n_f \right)  & \nonumber\\ 
 &\qquad+\alvp^3\,\left( 2368.193775 - 340.0691069\,n_f - 3.705612526\,\fl_{11}\,n_f + 9.472190428\,{n_f^2} \right.& \nonumber\\
&\left.\qquad+\fl_{02}\,\left( -24.01322539\,n_f + 0.7652692585\,{n_f^2} \right)  \right) 
& \nonumber \\
C^{\psi}_{L,8} & = 
0.5925925926\,\alvp + {\alvp^2}\,\left( 35.87664406 - 2.231471683\,n_f - 0.1217397796\,\fl_{02}\,n_f \right)  & \nonumber\\ 
 &\qquad+\alvp^3\,\left( 2215.210875 - 305.4730329\,n_f - 2.913702563\,\fl_{11}\,n_f + 8.337149534\,{n_f^2} \right.& \nonumber\\
&\left.\qquad+\fl_{02}\,\left( -12.97185267\,n_f + 0.344362391\,{n_f^2} \right)  \right) 
& \nonumber \\
C^{\psi}_{L,10} & = 
0.4848484848\,\alvp + {\alvp^2}\,\left( 32.01765947 - 1.908598248\,n_f - 0.06856377261\,\fl_{02}\,n_f \right)  & \nonumber\\ 
 &\qquad+\alvp^3\,\left( 2081.213222 - 278.0172177\,n_f - 2.397641695\,\fl_{11}\,n_f + 7.452505612\,{n_f^2} \right.& \nonumber\\
&\left.\qquad+\fl_{02}\,\left( -7.947555677\,n_f + 0.1841598535\,{n_f^2} \right)  \right) 
& \nonumber \\
C^{\psi}_{L,12} & = 
0.4102564103\,\alvp + {\alvp^2}\,\left( 29.0058065 - 1.671007435\,n_f - 0.04265241396\,\fl_{02}\,n_f \right)  & \nonumber\\ 
 &\qquad+\alvp^3\,\left( 1965.791047 - 255.8431044\,n_f - 2.035689631\,\fl_{11}\,n_f + 6.751061503\,{n_f^2} \right.& \nonumber\\
&\left.\qquad+\fl_{02}\,\left( -5.284573837\,n_f + 0.1098463658\,{n_f^2} \right)  \right) 
& \nonumber \\
C^{\psi,NS}_{L,14} & = 
0.3555555556\,\alvp + {\alvp^2}\,\left( 26.5848844 - 1.488624298\,n_f \right)  & \nonumber\\ 
 &\qquad+\alvp^3\,\left( 1866.009187 - 237.5642566\,n_f - 1.768138102\,\fl_{11}\,n_f + 6.182499654\,{n_f^2} \right)
& 
 \end{flalign*}
\begin{flalign*}
C^{G}_{L,2} & = 
0.6666666667\,\alvp\,n_f + 12.94776709\,{\alvp^2}\,n_f & \nonumber\\ 
 &\qquad+\alvp^3\,\left( 407.280632\,n_f - 20.23959748\,{n_f^2} - 0.388939664\,\fl^g_{11}\,{n_f^2} \right) 
& \nonumber \\
C^{G}_{L,4} & = 
0.2666666667\,\alvp\,n_f + 13.81659259\,{\alvp^2}\,n_f & \nonumber\\ 
 &\qquad+\alvp^3\,\left( 767.7125421\,n_f - 36.78419232\,{n_f^2} - 0.3984298404\,\fl^g_{11}\,{n_f^2} \right) 
& \nonumber \\
C^{G}_{L,6} & = 
0.1428571429\,\alvp\,n_f + 10.26095364\,{\alvp^2}\,n_f & \nonumber\\ 
 &\qquad+\alvp^3\,\left( 694.5092121\,n_f - 27.9895081\,{n_f^2} - 0.3055276256\,\fl^g_{11}\,{n_f^2} \right) 
& \nonumber \\
C^{G}_{L,8} & = 
0.08888888889\,\alvp\,n_f + 7.733545104\,{\alvp^2}\,n_f & \nonumber\\ 
 &\qquad+\alvp^3\,\left( 592.3307972\,n_f - 21.30333681\,{n_f^2} - 0.2322211886\,\fl^g_{11}\,{n_f^2} \right) 
& \nonumber \\
C^{G}_{L,10} & = 
0.06060606061\,\alvp\,n_f + 6.023053074\,{\alvp^2}\,n_f & \nonumber\\ 
 &\qquad+\alvp^3\,\left( 504.5424832\,n_f - 16.70544537\,{n_f^2} - 0.1805482229\,\fl^g_{11}\,{n_f^2} \right) 
& \nonumber \\
C^{G}_{L,12} & = 
0.04395604396\,\alvp\,n_f + 4.830676204\,{\alvp^2}\,n_f & \nonumber\\ 
 &\qquad+\alvp^3\,\left( 433.9050534\,n_f - 13.47418408\,{n_f^2} - 0.1439099345\,\fl^g_{11}\,{n_f^2} \right) 
& 
 \end{flalign*}

The numerical values of the anomalous dimensions for the odd moments
of $F_3$ read:

\begin{flalign*}
\gamma^{\mathrm{ns}}_{1} & = 
0
& \nonumber \\
\gamma^{\mathrm{ns}}_{3} & = 
5.555555556\,\alvp + {\alvp^2}\,\left( 70.88477366 - 5.12345679\,n_f \right)  & \nonumber\\ 
 &\qquad+\alvp^3\,\left( 1244.913602 - 196.4738081\,n_f - 1.762002743\,{n_f^2} \right) 
& \nonumber \\
\gamma^{\mathrm{ns}}_{5} & = 
8.088888889\,\alvp + {\alvp^2}\,\left( 98.19940741 - 7.68691358\,n_f \right)  & \nonumber\\ 
 &\qquad+\alvp^3\,\left( 1720.942172 - 278.1581739\,n_f - 2.366211248\,{n_f^2} \right) 
& \nonumber \\
\gamma^{\mathrm{ns}}_{7} & = 
9.780952381\,\alvp + {\alvp^2}\,\left( 116.4158903 - 9.437457798\,n_f \right)  & \nonumber\\ 
 &\qquad+\alvp^3\,\left( 2036.492478 - 330.8816595\,n_f - 2.739358023\,{n_f^2} \right) 
& \nonumber \\
\gamma^{\mathrm{ns}}_{9} & = 
11.05820106\,\alvp + {\alvp^2}\,\left( 130.3414045 - 10.77682428\,n_f \right)  & \nonumber\\ 
 &\qquad+\alvp^3\,\left( 2277.19805 - 370.5905277\,n_f - 3.006446616\,{n_f^2} \right) 
& \nonumber \\
\gamma^{\mathrm{ns}}_{11} & = 
12.08581049\,\alvp + {\alvp^2}\,\left( 141.6907901 - 11.86441897\,n_f \right)  & \nonumber\\ 
 &\qquad+\alvp^3\,\left( 2473.311857 - 402.691565\,n_f - 3.212753995\,{n_f^2} \right) 
& \nonumber \\
\gamma^{\mathrm{ns}}_{13} & = 
12.94606135\,\alvp + {\alvp^2}\,\left( 151.2989044 - 12.78102552\,n_f \right)  & \nonumber\\ 
 &\qquad+\alvp^3\,\left( 2639.409887 - 429.7314605\,n_f - 3.37996738\,{n_f^2} \right) 
& 
 \end{flalign*}

and the coefficient functions are:

\begin{flalign*}
C^{\mathrm{ns}}_{3,1} & = 
1 - 4\,\alvp + {\alvp^2}\,\left( -73.33333333 + 5.333333333\,n_f \right)  & \nonumber\\ 
 &\qquad+\alvp^3\,\left( -2652.154437 + 513.3100408\,n_f - 11.35802469\,{n_f^2} \right) 
& \nonumber \\
C^{\mathrm{ns}}_{3,3} & = 
1 + 1.666666667\,\alvp + {\alvp^2}\,\left( 14.25404015 - 6.742283951\,n_f \right)  & \nonumber\\ 
 &\qquad+\alvp^3\,\left( -839.7638717 - 45.09953407\,n_f + 1.747689309\,{n_f^2} \right) 
& \nonumber \\
C^{\mathrm{ns}}_{3,5} & = 
1 + 7.748148148\,\alvp + {\alvp^2}\,\left( 173.000629 - 19.39801646\,n_f \right)  & \nonumber\\ 
 &\qquad+\alvp^3\,\left( 4341.081057 - 961.2756356\,n_f + 22.24125078\,{n_f^2} \right) 
& \nonumber \\
C^{\mathrm{ns}}_{3,7} & = 
1 + 12.72248677\,\alvp + {\alvp^2}\,\left( 345.9910777 - 30.52332666\,n_f \right)  & \nonumber\\ 
 &\qquad+\alvp^3\,\left( 11119.00053 - 1960.237096\,n_f + 43.10377964\,{n_f^2} \right) 
& \nonumber \\
C^{\mathrm{ns}}_{3,9} & = 
1 + 16.9152381\,\alvp + {\alvp^2}\,\left( 520.0059615 - 40.35464229\,n_f \right)  & \nonumber\\ 
 &\qquad+\alvp^3\,\left( 18771.99642 - 2975.924131\,n_f + 63.17127673\,{n_f^2} \right) 
& \nonumber \\
C^{\mathrm{ns}}_{3,11} & = 
1 + 20.54831329\,\alvp + {\alvp^2}\,\left( 690.8719666 - 49.17096968\,n_f \right)  & \nonumber\\ 
 &\qquad+\alvp^3\,\left( 26941.47987 - 3984.411605\,n_f + 82.24581704\,{n_f^2} \right) 
& \nonumber \\
C^{\mathrm{ns}}_{3,13} & = 
1 + 23.76237745\,\alvp + {\alvp^2}\,\left( 857.1778817 - 57.18099124\,n_f \right)  & \nonumber\\ 
 &\qquad+\alvp^3\,\left( 35426.82868 - 4976.080869\,n_f + 100.3509187\,{n_f^2} \right) 
& 
 \end{flalign*}

\section{Acknowledgements}

A.R. would like to thank Sven Moch, Timo van Ritbergen and Thomas
Gehrmann for many useful discussions and Denny Fliegner for technical
support. J.V. would like to thank the University of Karlsruhe and the
DFG for repeated hospitality.

This work was supported by the DFG under Contract Ku 502/8-1 ({\it
DFG-Forschergruppe ``Quantenfeldtheorie, Computeralgebra und
Monte-Carlo-Simu\-lationen'' }) and the {\it Graduiertenkolleg
``Elementarteilchenphysik an Beschleunigern'' at the Univerit\"at
Karlsruhe}.

\begin{appendix}


\section{Conventions}

Here we give the complete expressions for the newly computed moments
and coefficient functions.

The notation of the color factors is as usual: The Casimir operators
of the fundamental and adjoint representation are denoted by $C_F$ and
$C_A$ and their values for the color group SU$(3)$ are $\frac{4}{3}$
and $3$, respectively. Additionally we are using the symmetric
structure constants of SU$(3)$ for which
$d^{abc}d^{abc}=\frac{40}{3}$. For the trace normalization of the
fundamental representation we have inserted $T_F = \frac{1}{2}$. For a
generic SU$(n)$ group the number of generators equals $N_A=n^2-1$.

The values of the Riemann $\zeta$ function are written as $\zeta_n =
\zeta(n)$. The only check for the newly computed 3-loop results is
from a large $n_f$-expansion of \cite{hep-ph/9401214} where the
$n_f^2$ terms are calculated. These terms are in agreement with ours.
To test the new parallel version of FORM we have also recalculated the
lower moments for $F_2$ and $F_L$ and found complete agreement with
\cite{nupha:b492:338} except for one missing term. In $C^G_{2,2}$
there should be the extra term $a_s^3 n_f C_F^2 (-\frac{160}{3}
\zeta_5)$. The numerical values given in this reference are correct.

It should be noted that the terms in $d^{abc}d^{abc}$ enter for the first 
time at the three loop level and help in the determination of 
$P^S_{qq}-P^S_{q\overline{q}}$.

\section{Results for the moments of $F_2$ and $F_L$}

\begin{eqnarray}
\label{gammaGG10}
\gamma^{GG}_{10} & = & 
 \alvp C_A\;
\left[
    + \frac{18421}{2310} 
\right] 
+ \;     \alvp     n_f\;
\left[
    + \frac{2}{3} 
\right]\Break
+ \;     \alvp^{2}     C_A^{2}\;
\left[
    + \frac{339202487377}{12326391000} 
\right] 
+ \;     \alvp^{2}     C_F     n_f\;
\left[
    + \frac{9284182}{4492125} 
\right] 
+ \;     \alvp^{2}     C_A     n_f\;
\left[
    - \frac{17481908}{1715175} 
\right]\Break
+ \;     \alvp^{3}     C_A^{3}\;
\left[
    + \frac{239083526238286750523}{1578596520362400000} 
\right]\Break
+ \;     \alvp^{3}     C_F     C_A     n_f\;
\left[
    - \frac{3374081335517123191}{36243287457300000} 
    + \frac{17831164}{190575}      \zeta_3
\right]\Break
+ \;     \alvp^{3}     C_F^{2}     n_f\;
\left[
    - \frac{3009386129483453}{3883209370425000} 
    - \frac{1344}{3025}      \zeta_3
\right]\Break
+ \;     \alvp^{3}     C_A^{2}     n_f\;
\left[
    + \frac{43228502203851731}{2196562876200000} 
    - \frac{17746492}{190575}      \zeta_3
\right]\Break
+ \;     \alvp^{3}     C_F     n_f^{2}\;
\left[
    - \frac{453912946493}{420260754375} 
\right] 
+ \;     \alvp^{3}     C_A     n_f^{2}\;
\left[
    - \frac{2752314359}{815051160} 
\right]\nonumber
{}
\end{eqnarray}
\begin{eqnarray}
\label{gammaGG12}
\gamma^{GG}_{12} & = & 
 \alvp C_A\;
\left[
    + \frac{71203}{8190} 
\right] 
+ \;     \alvp     n_f\;
\left[
    + \frac{2}{3} 
\right] 
+ \;     \alvp^{2}     C_A^{2}\;
\left[
    + \frac{16519839244157}{549353259000} 
\right]\Break
+ \;     \alvp^{2}     C_A     n_f\;
\left[
    - \frac{23220103}{2108106} 
\right] 
+ \;     \alvp^{2}     C_F     n_f\;
\left[
    + \frac{74823503}{36540504} 
\right]\Break
+ \;     \alvp^{3}     C_A^{3}\;
\left[
    + \frac{6993119873800651614841}{42112541869725600000} 
\right]\Break
+ \;     \alvp^{3}     C_F     C_A     n_f\;
\left[
    - \frac{49693541388602890695713}{498238162032042432000} 
    + \frac{58085396}{585585}      \zeta_3
\right]\Break
+ \;     \alvp^{3}     C_F^{2}     n_f\;
\left[
    - \frac{4699124115250144376149}{5480619782352466752000} 
    - \frac{158}{507}      \zeta_3
\right]\Break
+ \;     \alvp^{3}     C_A^{2}     n_f\;
\left[
    + \frac{74338654569222233539}{3871314390303360000} 
    - \frac{57902906}{585585}      \zeta_3
\right]\Break
+ \;     \alvp^{3}     C_F     n_f^{2}\;
\left[
    - \frac{50627726543561953}{45626205314289600} 
\right]\Break
+ \;     \alvp^{3}     C_A     n_f^{2}\;
\left[
    - \frac{2635361358193}{759677078160} 
\right]\nonumber
{}
\end{eqnarray}
\begin{eqnarray}
\label{gammaGQ10}
\gamma^{G\psi}_{10} & = & 
 \alvp C_F\;
\left[
    - \frac{112}{495} 
\right]\Break
+ \;     \alvp^{2}     C_F     C_A\;
\left[
    - \frac{133349533}{80858250} 
\right] 
+ \;     \alvp^{2}     C_F^{2}\;
\left[
    + \frac{88631998}{121287375} 
\right] 
+ \;     \alvp^{2}     C_F     n_f\;
\left[
    - \frac{74368}{735075} 
\right]\Break
+ \;     \alvp^{3}     C_F     C_A^{2}\;
\left[
    - \frac{165894431274725803}{48324383276400000} 
    - \frac{645584}{81675}      \zeta_3
\right]\Break
+ \;     \alvp^{3}     C_F^{2}     C_A\;
\left[
    - \frac{4224421791031474951}{217459724743800000} 
    + \frac{645584}{27225}      \zeta_3
\right]\Break
+ \;     \alvp^{3}     C_F^{3}\;
\left[
    + \frac{6457897459084371893}{326189587115700000} 
    - \frac{1291168}{81675}      \zeta_3
\right]\Break
+ \;     \alvp^{3}     C_F     C_A     n_f\;
\left[
    - \frac{18846629176433}{47069204490000} 
    + \frac{1792}{495}      \zeta_3
\right]\Break
+ \;     \alvp^{3}     C_F^{2}     n_f\;
\left[
    + \frac{529979902254031}{1294403123475000} 
    - \frac{1792}{495}      \zeta_3
\right] 
+ \;     \alvp^{3}     C_F     n_f^{2}\;
\left[
    + \frac{152267426}{363862125} 
\right]\nonumber
{}
\end{eqnarray}
\begin{eqnarray}
\label{gammaGQ12}
\gamma^{G\psi}_{12} & = & 
 \alvp C_F\;
\left[
    - \frac{79}{429} 
\right] 
+ \;     \alvp^{2}     C_F     C_A\;
\left[
    - \frac{70863259553}{50645138544} 
\right] 
+ \;     \alvp^{2}     C_F^{2}\;
\left[
    + \frac{9387059226553}{13927413099600} 
\right]\Break
+ \;     \alvp^{2}     C_F     n_f\;
\left[
    - \frac{14257247}{115945830} 
\right]\Break
+ \;     \alvp^{3}     C_F     C_A^{2}\;
\left[
    - \frac{24751969909767240119153}{14947144860961272960000} 
    - \frac{46179137}{6441435}      \zeta_3
\right]\Break
+ \;     \alvp^{3}     C_F^{2}     C_A\;
\left[
    - \frac{115986599183122809974023}{5872092623949071520000} 
    + \frac{46179137}{2147145}      \zeta_3
\right]\Break
+ \;     \alvp^{3}     C_F^{3}\;
\left[
    + \frac{2165927305724992215894703}{113037783011019626760000} 
    - \frac{92358274}{6441435}      \zeta_3
\right]\Break
+ \;     \alvp^{3}     C_F     C_A     n_f\;
\left[
    - \frac{64190493078139789}{195540879918384000} 
    + \frac{1264}{429}      \zeta_3
\right]\Break
+ \;     \alvp^{3}     C_F^{2}     n_f\;
\left[
    + \frac{1401404001326440151}{13981172914164456000} 
    - \frac{1264}{429}      \zeta_3
\right]\Break
+ \;     \alvp^{3}     C_F     n_f^{2}\;
\left[
    + \frac{13454024393417}{41782239298800} 
\right]\nonumber
{}
\end{eqnarray}
\begin{eqnarray}
\label{gammaQG10}
\gamma^{\psi G}_{10} & = & 
 \alvp n_f\;
\left[
    - \frac{28}{165} 
\right] 
+ \;     \alvp^{2}     C_F     n_f\;
\left[
    - \frac{379479917}{125779500} 
\right] 
+ \;     \alvp^{2}     C_A     n_f\;
\left[
    + \frac{373810079}{150935400} 
\right]\Break
+ \;     \alvp^{3}     C_F     C_A     n_f\;
\left[
    + \frac{926990216580622991}{24162191638200000} 
    - \frac{643396}{27225}      \zeta_3
\right]\Break
+ \;     \alvp^{3}     C_F^{2}     n_f\;
\left[
    - \frac{1091980048536213833}{54364931185950000} 
    + \frac{17712}{3025}      \zeta_3
\right]\Break
+ \;     \alvp^{3}     C_A^{2}     n_f\;
\left[
    - \frac{21025430857658971}{1022738270400000} 
    + \frac{483988}{27225}      \zeta_3
\right]\Break
+ \;     \alvp^{3}     C_F     n_f^{2}\;
\left[
    + \frac{1584713325754369}{2588806246950000} 
\right] 
+ \;     \alvp^{3}     C_A     n_f^{2}\;
\left[
    + \frac{1669885489}{7906140000} 
\right]\nonumber
{}
\end{eqnarray}
\begin{eqnarray}
\label{gammaQG12}
\gamma^{\psi G}_{12} & = & 
 \alvp n_f\;
\left[
    - \frac{79}{546} 
\right]\Break
+ \;     \alvp^{2}     C_F     n_f\;
\left[
    - \frac{9256843807}{3197294100} 
\right] 
+ \;     \alvp^{2}     C_A     n_f\;
\left[
    + \frac{653436358741}{268572704400} 
\right]\Break
+ \;     \alvp^{3}     C_F     C_A     n_f\;
\left[
    + \frac{4046032530021008148641959}{104630014026728910720000} 
    - \frac{171207527}{8198190}      \zeta_3
\right]\Break
+ \;     \alvp^{3}     C_F^{2}     n_f\;
\left[
    - \frac{2960118366121154186145047}{143866269286752252240000} 
    + \frac{2563}{507}      \zeta_3
\right]\Break
+ \;     \alvp^{3}     C_A^{2}     n_f\;
\left[
    - \frac{10876559659107463644949}{543532540398591744000} 
    + \frac{129763817}{8198190}      \zeta_3
\right]\Break
+ \;     \alvp^{3}     C_F     n_f^{2}\;
\left[
    + \frac{149081947693135635881}{249119081016021216000} 
\right]\Break
+ \;     \alvp^{3}     C_A     n_f^{2}\;
\left[
    + \frac{226617401255197}{4399220898072000} 
\right]\nonumber
{}
\end{eqnarray}
\begin{eqnarray}
\label{gammaQQ10}
\gamma^{\psi\psi}_{10} & = & 
 \alvp C_F\;
\left[
    + \frac{12055}{1386} 
\right]\Break
+ \;     \alvp^{2}     C_F     C_A\;
\left[
    + \frac{19524247733}{523908000} 
\right] 
+ \;     \alvp^{2}     C_F^{2}\;
\left[
    - \frac{9579051036701}{1331250228000} 
\right]\Break
+ \;     \alvp^{2}     C_F     n_f\;
\left[
    - \frac{2451995507}{288149400} 
\right] 
+ \;     \alvp^{2}     \fl_{02}     C_F     n_f\;
\left[
    - \frac{27284}{13476375} 
\right]\Break
+ \;     \alvp^{3}     C_F     C_A^{2}\;
\left[
    + \frac{94091568579766453}{435681892800000} 
    + \frac{151796299}{8004150}      \zeta_3
\right]\Break
+ \;     \alvp^{3}     C_F^{2}     C_A\;
\left[
    - \frac{16389982059548833}{465937579800000} 
    - \frac{151796299}{2668050}      \zeta_3
\right]\Break
+ \;     \alvp^{3}     C_F^{3}\;
\left[
    - \frac{2207711300808736405687}{127866318149354400000} 
    + \frac{151796299}{4002075}      \zeta_3
\right]\Break
+ \;     \alvp^{3}     C_F     C_A     n_f\;
\left[
    - \frac{9007773127403}{389001690000} 
    - \frac{48220}{693}      \zeta_3
\right]\Break
+ \;     \alvp^{3}     C_F^{2}     n_f\;
\left[
    - \frac{75522073210471127}{1230075210672000} 
    + \frac{48220}{693}      \zeta_3
\right]\Break
+ \;     \alvp^{3}     C_F     n_f^{2}\;
\left[
    - \frac{27995901056887}{11981252052000} 
\right]\Break
+ \;     \alvp^{3}     \fl_{02}     C_F     C_A     n_f\;
\left[
    - \frac{1028766412107043}{5177612493900000} 
    - \frac{12544}{27225}      \zeta_3
\right]\Break
+ \;     \alvp^{3}     \fl_{02}     C_F^{2}     n_f\;
\left[
    + \frac{209966063746798}{485401171303125} 
    + \frac{12544}{27225}      \zeta_3
\right]\Break
+ \;     \alvp^{3}     \fl_{02}     C_F     n_f^{2}\;
\left[
    - \frac{33230913134}{420260754375} 
\right]\nonumber
{}
\end{eqnarray}
\begin{eqnarray}
\label{gammaQQ12}
\gamma^{\psi\psi}_{12} & = & 
 \alvp C_F\;
\left[
    + \frac{423424}{45045} 
\right]\Break
+ \;     \alvp^{2}     C_F     C_A\;
\left[
    + \frac{19487270392267}{486972486000} 
\right] 
+ \;     \alvp^{2}     C_F^{2}\;
\left[
    - \frac{5507868301548461}{731189187729000} 
\right]\Break
+ \;     \alvp^{2}     C_F     n_f\;
\left[
    - \frac{90143221429}{9739449720} 
\right] 
+ \;     \alvp^{2}     \fl_{02}     C_F     n_f\;
\left[
    - \frac{249775}{255783528} 
\right]\Break
+ \;     \alvp^{3}     C_F     C_A^{2}\;
\left[
    + \frac{1395004186652448755863}{6016642459027200000} 
    + \frac{25648239313}{1352701350}      \zeta_3
\right]\Break
+ \;     \alvp^{3}     C_F^{2}     C_A\;
\left[
    - \frac{98204412073910020058227}{2634913356900224400000} 
    - \frac{25648239313}{450900450}      \zeta_3
\right]\Break
+ \;     \alvp^{3}     C_F^{3}\;
\left[
    - \frac{81630141011772791446330057}{4747586886462824323920000} 
    + \frac{25648239313}{676350675}      \zeta_3
\right]\Break
+ \;     \alvp^{3}     C_F     C_A     n_f\;
\left[
    - \frac{25478252190337435009}{1052912430329760000} 
    - \frac{3387392}{45045}      \zeta_3
\right]\Break
+ \;     \alvp^{3}     C_F^{2}     n_f\;
\left[
    - \frac{35346062280941906036867}{526982671380044880000} 
    + \frac{3387392}{45045}      \zeta_3
\right]\Break
+ \;     \alvp^{3}     C_F     n_f^{2}\;
\left[
    - \frac{65155853387858071}{26322810758244000} 
\right]\Break
+ \;     \alvp^{3}     \fl_{02}     C_F     C_A     n_f\;
\left[
    - \frac{69697489543846494691}{332158774688028288000} 
    - \frac{12482}{39039}      \zeta_3
\right]\Break
+ \;     \alvp^{3}     \fl_{02}     C_F^{2}     n_f\;
\left[
    + \frac{86033255402443256197}{219224791294098670080} 
    + \frac{12482}{39039}      \zeta_3
\right]\Break
+ \;     \alvp^{3}     \fl_{02}     C_F     n_f^{2}\;
\left[
    - \frac{2566080055386457}{45626205314289600} 
\right]\nonumber
{}
\end{eqnarray}
\begin{eqnarray}
\label{gammaQQ14}
\gamma^{\mathrm{ns}}_{14} & = & 
 a_s C_F\;
\left[
    + \frac{180121}{18018} 
\right]\Break
+ \;     a_s^{2}     C_F     C_A\;
\left[
    + \frac{288858136265399}{6817614804000} 
\right]\Break
+ \;     a_s^{2}     C_F     n_f\;
\left[
    - \frac{481761665447}{48697248600} 
\right]\Break
+ \;     a_s^{2}     C_F^{2}\;
\left[
    - \frac{22819142381313407}{2924756750916000} 
\right]\Break
+ \;     a_s^{3}     C_F     C_A     n_f\;
\left[
    - \frac{92531316363319241549}{3685193506154160000} 
    - \frac{720484}{9009}      \zeta_3
\right]\Break
+ \;     a_s^{3}     C_F     C_A^{2}\;
\left[
    + \frac{126653245164236390142889}{515927090861582400000} 
    + \frac{3663695353}{193243050}      \zeta_3
\right]\Break
+ \;     a_s^{3}     C_F     n_f^{2}\;
\left[
    - \frac{68167166257767019}{26322810758244000} 
\right]\Break
+ \;     a_s^{3}     C_F^{2}     C_A\;
\left[
    - \frac{96001333716903621488387}{2459252466440209440000} 
    - \frac{3663695353}{64414350}      \zeta_3
\right]\Break
+ \;     a_s^{3}     C_F^{2}     n_f\;
\left[
    - \frac{37908544797975614512733}{526982671380044880000} 
    + \frac{720484}{9009}      \zeta_3
\right]\Break
+ \;     a_s^{3}     C_F^{3}\;
\left[
    - \frac{40552395746064871242211709}{2373793443231412161960000} 
    + \frac{3663695353}{96621525}      \zeta_3
\right]\nonumber
{}
\end{eqnarray}

\begin{eqnarray}
\label{C2Q10}
C^{\psi}_{2,10} & = & 
1\;
{}
+ \; \alvp C_F\;
\left[
    + \frac{2006299}{138600} 
\right] 
+ \;     \alvp^{2}     C_F     C_A\;
\left[
    + \frac{6124093193824187}{29045459520000} 
    - \frac{104674}{1155}      \zeta_3
\right]\Break
+ \;     \alvp^{2}     C_F     n_f\;
\left[
    - \frac{561457267429757}{15975002736000} 
\right] 
+ \;     \alvp^{2}     C_F^{2}\;
\left[
    + \frac{558708799987324013}{14760902528064000} 
    + \frac{88798}{1155}      \zeta_3
\right]\Break
+ \;     \alvp^{2}     \fl_{02}     C_F     n_f\;
\left[
    + \frac{3584203788491}{15689734830000} 
\right]\Break
+ \;     \alvp^{3}     C_F     C_A     n_f\;
\left[
    - \frac{21664244926039357214987}{23550349033411200000} 
    + \frac{10519793104}{42567525}      \zeta_3
    - \frac{24110}{693}      \zeta_4
\right]\Break
+ \;     \alvp^{3}     C_F     C_A^{2}\;
\left[
    + \frac{709221119965457939095237}{235503490334112000000} 
    - \frac{14713925739913}{6243237000}      \zeta_3
\right. \Break \left. \qquad \qquad
    + \frac{151796299}{16008300}      \zeta_4
    + \frac{190858}{231}      \zeta_5
\right]\Break
+ \;     \alvp^{3}     C_F     n_f^{2}\;
\left[
    + \frac{57084428047851551911}{996360920644320000} 
    + \frac{48220}{18711}      \zeta_3
\right]\Break
+ \;     \alvp^{3}     C_F^{2}     C_A\;
\left[
    + \frac{16350009304926933389608829}{8369431733412288000000} 
    + \frac{1430215936081}{6163195500}      \zeta_3
\right. \Break \left. \qquad \qquad
    - \frac{151796299}{5336100}      \zeta_4
    - \frac{22658}{99}      \zeta_5
\right]\Break
+ \;     \alvp^{3}     C_F^{2}     n_f\;
\left[
    - \frac{1521387460036994061010049}{2720065313358993600000} 
    - \frac{3997754476}{42567525}      \zeta_3
    + \frac{24110}{693}      \zeta_4
\right]\Break
+ \;     \alvp^{3}     C_F^{3}\;
\left[
    - \frac{3247779532370920623770610131}{92155812816602703168000000} 
    + \frac{2182208825245282}{1622461215375}      \zeta_3
\right. \Break \left. \qquad \qquad
    + \frac{151796299}{8004150}      \zeta_4
    - \frac{75212}{99}      \zeta_5
\right]\Break
+ \;     \alvp^{3}     \fl_{02}     C_F     C_A     n_f\;
\left[
    + \frac{3566946294536415188593}{861140509985448000000} 
    - \frac{79527463}{56600775}      \zeta_3
    - \frac{6272}{27225}      \zeta_4
\right]\Break
+ \;     \alvp^{3}     \fl_{02}     C_F     n_f^{2}\;
\left[
    - \frac{148475806971656561}{244642190336775000} 
    + \frac{33008}{735075}      \zeta_3
\right]\Break
+ \;     \alvp^{3}     \fl_{02}     C_F^{2}     n_f\;
\left[
    + \frac{422577250875954453771617}{58772839806506826000000} 
    - \frac{12949105012}{11037151125}      \zeta_3
    + \frac{6272}{27225}      \zeta_4
\right]\Break
+ \;     \alvp^{3}     \fl_{11}     n_f     { \frac{d^{abc}d^{abc}}{n} }\;
\left[
    + \frac{3753913187503}{352066176000} 
    + \frac{81388}{606375}      \zeta_3
    - \frac{448}{33}      \zeta_5
\right]\nonumber
{}
\end{eqnarray}
\begin{eqnarray}
\label{C2Q12}
C^{\psi}_{2,12} & = & 
1\;
{}
+ \; \alvp C_F\;
\left[
    + \frac{183473419}{10810800} 
\right]\Break
+ \;     \alvp^{2}     C_F     C_A\;
\left[
    + \frac{21388499873332252399}{87742702527480000} 
    - \frac{1477711}{15015}      \zeta_3
\right]\Break
+ \;     \alvp^{2}     C_F^{2}\;
\left[
    + \frac{9127110915702407798941}{131745667845011220000} 
    + \frac{1261726}{15015}      \zeta_3
\right]\Break
+ \;     \alvp^{2}     C_F     n_f\;
\left[
    - \frac{57904356630607013}{1403883240439680} 
\right] 
+ \;     \alvp^{2}     \fl_{02}     C_F     n_f\;
\left[
    + \frac{103555928663269}{563286485361600} 
\right]\Break
+ \;     \alvp^{3}     C_F     C_A     n_f\;
\left[
    - \frac{83712626229337204073275967}{75885504678726462720000} 
    + \frac{36842282041}{133783650}      \zeta_3
\right. \Break \left. \qquad \qquad
    - \frac{1693696}{45045}      \zeta_4
\right] 
+ \;     \alvp^{3}     C_F     C_A^{2}\;
\left[
    + \frac{163181620367687907864404054279}{45531302807235877632000000} 
\right. \Break \left. \qquad \qquad
    - \frac{641004330821357}{243486243000}      \zeta_3
    + \frac{25648239313}{2705402700}      \zeta_4
    + \frac{2642336}{3003}      \zeta_5
\right]\Break
+ \;     \alvp^{3}     C_F     n_f^{2}\;
\left[
    + \frac{2003755100099657438423887}{28457064254522423520000} 
    + \frac{3387392}{1216215}      \zeta_3
\right]\Break
+ \;     \alvp^{3}     C_F^{2}     C_A\;
\left[
    + \frac{4843191986526849157300333804109}{1709131279126616756611200000} 
    + \frac{4948562279669051}{97491891697200}      \zeta_3
\right. \Break \left. \qquad \qquad
    - \frac{25648239313}{901800900}      \zeta_4
    - \frac{241094}{1287}      \zeta_5
\right]\Break
+ \;     \alvp^{3}     C_F^{2}     n_f\;
\left[
    - \frac{848389810670975600831798557}{1130004151488672235776000} 
    - \frac{1326399435709}{12174312150}      \zeta_3
\right. \Break \left. \qquad \qquad
    + \frac{1693696}{45045}      \zeta_4
\right] 
+ \;     \alvp^{3}     C_F^{3}\;
\left[
    + \frac{131770367393773533536363889790633}{3421680820811486746735622400000} 
\right. \Break \left. \qquad \qquad
    + \frac{42634681331415644}{24926904127125}      \zeta_3
    + \frac{25648239313}{1352701350}      \zeta_4
    - \frac{88004}{99}      \zeta_5
\right]\Break
+ \;     \alvp^{3}     \fl_{02}     C_F     C_A     n_f\;
\left[
    + \frac{86565903158314138806418902631}{26931765610480021619328000000} 
    - \frac{3109476222803}{3165321159000}      \zeta_3
\right. \Break \left. \qquad \qquad
    - \frac{6241}{39039}      \zeta_4
\right] 
+ \;     \alvp^{3}     \fl_{02}     C_F     n_f^{2}\;
\left[
    - \frac{1960867603733060624851}{4404069467961803640000} 
    + \frac{2780}{95823}      \zeta_3
\right]\Break
+ \;     \alvp^{3}     \fl_{02}     C_F^{2}     n_f\;
\left[
    + \frac{33528586559068805780843402423}{5290168244915718532368000000} 
    - \frac{751723347541}{791330289750}      \zeta_3
\right. \Break \left. \qquad \qquad
    + \frac{6241}{39039}      \zeta_4
\right]\Break
+ \;     \alvp^{3}     \fl_{11}     n_f     { \frac{d^{abc}d^{abc}}{n} }\;
\left[
    + \frac{809917806143013559}{67966375276800000} 
    + \frac{622064791}{851350500}      \zeta_3
    - \frac{200}{13}      \zeta_5
\right]\nonumber
{}
\end{eqnarray}
\begin{eqnarray}
\label{C2Q14}
C^{\mathrm{ns}}_{2,14} & = & 
1\;
{}
+ \; a_s C_F\;
\left[
    + \frac{90849502}{4729725} 
\right]\Break
+ \;     a_s^{2}     C_F     C_A\;
\left[
    + \frac{1345455725874078602801}{4913591341538880000} 
    - \frac{315626}{3003}      \zeta_3
\right]\Break
+ \;     a_s^{2}     C_F     n_f\;
\left[
    - \frac{1644267296654871017}{35097081010992000} 
\right]\Break
+ \;     a_s^{2}     C_F^{2}\;
\left[
    + \frac{31002322638187643268973}{301132955074311360000} 
    + \frac{271010}{3003}      \zeta_3
\right]\Break
+ \;     a_s^{3}     C_F     C_A     n_f\;
\left[
    - \frac{28812973254576289068812626927}{22575937641921122659200000} 
    + \frac{31112773830559}{103481653275}      \zeta_3
\right. \Break \left. \qquad \qquad
    - \frac{360242}{9009}      \zeta_4
\right]\Break
+ \;     a_s^{3}     C_F     C_A^{2}\;
\left[
    + \frac{7823621156350047175125815627621}{1896378761921374303372800000} 
    - \frac{83168919211026563}{28974862917000}      \zeta_3
\right. \Break \left. \qquad \qquad
    + \frac{3663695353}{386486100}      \zeta_4
    + \frac{2738146}{3003}      \zeta_5
\right]\Break
+ \;     a_s^{3}     C_F     n_f^{2}\;
\left[
    + \frac{2361466163828853440218087}{28457064254522423520000} 
    + \frac{720484}{243243}      \zeta_3
\right]\Break
+ \;     a_s^{3}     C_F^{2}     C_A\;
\left[
    + \frac{64116556842577537005311631547547}{17091312791266167566112000000} 
    - \frac{13682992796062061}{81243243081000}      \zeta_3
\right. \Break \left. \qquad \qquad
    - \frac{3663695353}{128828700}      \zeta_4
    - \frac{865562}{9009}      \zeta_5
\right]\Break
+ \;     a_s^{3}     C_F^{2}     n_f\;
\left[
    - \frac{358451845381207175240043774137}{377340672014967335875200000} 
    - \frac{12775152582499}{103481653275}      \zeta_3
    + \frac{360242}{9009}      \zeta_4
\right]\Break
+ \;     a_s^{3}     C_F^{3}\;
\left[
    + \frac{12924407779985031268428280066600921}{72710717442244093368131976000000} 
    + \frac{3556746663996971701}{1695029480644500}      \zeta_3
\right. \Break \left. \qquad \qquad
    + \frac{3663695353}{193243050}      \zeta_4
    - \frac{9512108}{9009}      \zeta_5
\right]\Break
+ \;     a_s^{3}     fl_{11}     n_f     { \frac{d^{abc}d^{abc}}{n} }\;
\left[
    + \frac{637395762233410021}{49587712574400000} 
    + \frac{78376866703}{65553988500}      \zeta_3
    - \frac{352}{21}      \zeta_5
\right]\nonumber
{}
\end{eqnarray}
\begin{eqnarray}
\label{CLQ10}
C^{\psi}_{L,10} & = & 
 \alvp C_F\;
\left[
    + \frac{4}{11} 
\right]\Break
+ \;     \alvp^{2}     C_F     C_A\;
\left[
    + \frac{89670761}{8731800} 
    - \frac{48}{11}      \zeta_3
\right]\Break
+ \;     \alvp^{2}     C_F     n_f\;
\left[
    - \frac{163679}{114345} 
\right]\Break
+ \;     \alvp^{2}     C_F^{2}\;
\left[
    - \frac{1999510607}{528273900} 
    + \frac{96}{11}      \zeta_3
\right]\Break
+ \;     \alvp^{2}     \fl_{02}     C_F     n_f\;
\left[
    - \frac{415796}{8085825} 
\right]\Break
+ \;     \alvp^{3}     C_F     C_A     n_f\;
\left[
    - \frac{176183576988227323}{1699159381920000} 
    + \frac{55485434}{1216215}      \zeta_3
\right]\Break
+ \;     \alvp^{3}     C_F     C_A^{2}\;
\left[
    + \frac{2366034921481985137}{6796637527680000} 
    - \frac{95022195887}{187297110}      \zeta_3
    + \frac{3760}{11}      \zeta_5
\right]\Break
+ \;     \alvp^{3}     C_F     n_f^{2}\;
\left[
    + \frac{63272639}{11320155} 
\right]\Break
+ \;     \alvp^{3}     C_F^{2}     C_A\;
\left[
    - \frac{323139848004267269}{3354750574560000} 
    + \frac{22904191}{17325}      \zeta_3
    - \frac{14240}{11}      \zeta_5
\right]\Break
+ \;     \alvp^{3}     C_F^{2}     n_f\;
\left[
    + \frac{9048874326307637}{190368782604000} 
    - \frac{1174256}{15015}      \zeta_3
\right]\Break
+ \;     \alvp^{3}     C_F^{3}\;
\left[
    - \frac{887562386698645967383}{3166213592269728000} 
    - \frac{357031607224}{468242775}      \zeta_3
    + \frac{13440}{11}      \zeta_5
\right]\Break
+ \;     \alvp^{3}     \fl_{02}     C_F     C_A     n_f\;
\left[
    - \frac{68379915239899511}{43491944948760000} 
    - \frac{36224}{147015}      \zeta_3
\right]\Break
+ \;     \alvp^{3}     \fl_{02}     C_F     n_f^{2}\;
\left[
    + \frac{21670644503}{156897348300} 
\right]\Break
+ \;     \alvp^{3}     \fl_{02}     C_F^{2}     n_f\;
\left[
    - \frac{319520059852805113}{282697642166940000} 
    + \frac{1373248}{1911195}      \zeta_3
\right]\Break
+ \;     \alvp^{3}     \fl_{11}     n_f     { \frac{d^{abc}d^{abc}}{n} }\;
\left[
    - \frac{5073093424963}{528099264000} 
    - \frac{1820773}{363825}      \zeta_3
    + \frac{160}{11}      \zeta_5
\right]\nonumber
{}
\end{eqnarray}
\begin{eqnarray}
\label{CLQ12}
C^{\psi}_{L,12} & = & 
 \alvp C_F\;
\left[
    + \frac{4}{13} 
\right] 
+ \;     \alvp^{2}     C_F     C_A\;
\left[
    + \frac{7126442885209}{811620810000} 
    - \frac{48}{13}      \zeta_3
\right]\Break
+ \;     \alvp^{2}     C_F     n_f\;
\left[
    - \frac{2201663}{1756755} 
\right] 
+ \;     \alvp^{2}     C_F^{2}\;
\left[
    - \frac{12296141077867}{5275535265000} 
    + \frac{96}{13}      \zeta_3
\right]\Break
+ \;     \alvp^{2}     \fl_{02}     C_F     n_f\;
\left[
    - \frac{16072451}{502431930} 
\right] 
+ \;     \alvp^{3}     C_F     n_f^{2}\;
\left[
    + \frac{5203557911}{1027701675} 
\right]\Break
+ \;     \alvp^{3}     C_F     C_A     n_f\;
\left[
    - \frac{61335054761825219657}{658070268956100000} 
    + \frac{2151978544}{52026975}      \zeta_3
\right]\Break
+ \;     \alvp^{3}     C_F     C_A^{2}\;
\left[
    + \frac{14897818968827338964411}{52645621516488000000} 
    - \frac{5467652405147}{10145260125}      \zeta_3
    + \frac{5600}{13}      \zeta_5
\right]\Break
+ \;     \alvp^{3}     C_F^{2}     C_A\;
\left[
    + \frac{1508964059584735294818791}{26349133569002244000000} 
    + \frac{3065115509662}{2029052025}      \zeta_3
    - \frac{21600}{13}      \zeta_5
\right]\Break
+ \;     \alvp^{3}     C_F^{2}     n_f\;
\left[
    + \frac{12143703744427185053}{316848648015900000} 
    - \frac{47440688}{675675}      \zeta_3
\right]\Break
+ \;     \alvp^{3}     C_F^{3}\;
\left[
    - \frac{16091344678479668687707841}{42817342049628646500000} 
    - \frac{10180155542576}{10145260125}      \zeta_3
    + 1600     \zeta_5
\right]\Break
+ \;     \alvp^{3}     \fl_{02}     C_F     C_A     n_f\;
\left[
    - \frac{592278098533386728681}{576664539388938000000} 
    - \frac{561855512}{3381753375}      \zeta_3
\right]\Break
+ \;     \alvp^{3}     \fl_{02}     C_F     n_f^{2}\;
\left[
    + \frac{435058406339681}{5280810800265000} 
\right]\Break
+ \;     \alvp^{3}     \fl_{02}     C_F^{2}     n_f\;
\left[
    - \frac{65983065928499265747263}{85634684099257293000000} 
    + \frac{1570446544}{3381753375}      \zeta_3
\right]\Break
+ \;     \alvp^{3}     \fl_{11}     n_f     { \frac{d^{abc}d^{abc}}{n} }\;
\left[
    - \frac{503821438649257451}{62302510670400000} 
    - \frac{302982523}{70945875}      \zeta_3
    + \frac{160}{13}      \zeta_5
\right]\nonumber
{}
\end{eqnarray}
\begin{eqnarray}
\label{CLQ14}
C^{\mathrm{ns}}_{L,14} & = & 
 a_s C_F\;
\left[
    + \frac{4}{15} 
\right]\Break
+ \;     a_s^{2}     C_F     C_A\;
\left[
    + \frac{3736751546509}{486972486000} 
    - \frac{16}{5}      \zeta_3
\right]\Break
+ \;     a_s^{2}     C_F     n_f\;
\left[
    - \frac{2263109}{2027025} 
\right]\Break
+ \;     a_s^{2}     C_F^{2}\;
\left[
    - \frac{6706232197}{4969107000} 
    + \frac{32}{5}      \zeta_3
\right]\Break
+ \;     a_s^{3}     C_F     C_A     n_f\;
\left[
    - \frac{14479744644122508496943}{170858971648965600000} 
    + \frac{16947716309}{447972525}      \zeta_3
\right]\Break
+ \;     a_s^{3}     C_F     C_A^{2}\;
\left[
    + \frac{1704131316277822389591041}{7517794752554486400000} 
    - \frac{47503400282843}{82785322620}      \zeta_3
    + \frac{1552}{3}      \zeta_5
\right]\Break
+ \;     a_s^{3}     C_F     n_f^{2}\;
\left[
    + \frac{422957746}{91216125} 
\right]\Break
+ \;     a_s^{3}     C_F^{2}     C_A\;
\left[
    + \frac{1016513978878471248683819}{5269826713800448800000} 
    + \frac{51799344112951}{30435780375}      \zeta_3
    - 2016     \zeta_5
\right]\Break
+ \;     a_s^{3}     C_F^{2}     n_f\;
\left[
    + \frac{58345189864914275683}{1864532428708950000} 
    - \frac{438969448}{6891885}      \zeta_3
\right]\Break
+ \;     a_s^{3}     C_F^{3}\;
\left[
    - \frac{8328506007291432016890749}{17917410826921525920000} 
    - \frac{640509641943719}{517408266375}      \zeta_3
    + \frac{5888}{3}      \zeta_5
\right]\Break
+ \;     a_s^{3}     fl_{11}     n_f     { \frac{d^{abc}d^{abc}}{n} }\;
\left[
    - \frac{110339419075223314037}{15793686454946400000} 
    - \frac{17420356529}{4682427750}      \zeta_3
    + \frac{32}{3}      \zeta_5
\right]\nonumber
{}
\end{eqnarray}
\begin{eqnarray}
\label{C2G10}
C^{G}_{2,10} & = & 
 \alvp n_f\;
\left[
    - \frac{4352}{7425} 
\right]\Break
+ \;     \alvp^{2}     C_A     n_f\;
\left[
    - \frac{651112454591}{185952412800} 
    - \frac{54}{55}      \zeta_3
\right]\Break
+ \;     \alvp^{2}     C_F     n_f\;
\left[
    - \frac{72533010722807}{6973215480000} 
    + \frac{108}{55}      \zeta_3
\right]\Break
+ \;     \alvp^{3}     C_A     n_f^{2}\;
\left[
    + \frac{30367858943250477461}{1739677797950400000} 
    - \frac{18965}{137214}      \zeta_3
\right]\Break
+ \;     \alvp^{3}     C_A^{2}     n_f\;
\left[
    - \frac{114563089982050063118447}{669775952210904000000} 
    + \frac{427785744377}{7924108500}      \zeta_3
\right. \Break \left. \qquad \qquad
    + \frac{241994}{27225}      \zeta_4
    + \frac{8}{33}      \zeta_5
\right]\Break
+ \;     \alvp^{3}     C_F     C_A     n_f\;
\left[
    - \frac{110850413975318446061177}{2487739251069072000000} 
    + \frac{7772983651}{7358100750}      \zeta_3
\right. \Break \left. \qquad \qquad
    - \frac{321698}{27225}      \zeta_4
    + \frac{244}{11}      \zeta_5
\right]\Break
+ \;     \alvp^{3}     C_F     n_f^{2}\;
\left[
    + \frac{25164738348656825457229}{1399353328726353000000} 
    - \frac{2762978063}{1226350125}      \zeta_3
\right]\Break
+ \;     \alvp^{3}     C_F^{2}     n_f\;
\left[
    - \frac{26284376777719892724358177}{235091359226027304000000} 
    + \frac{1782724946402}{77260057875}      \zeta_3
\right. \Break \left. \qquad \qquad
    + \frac{8856}{3025}      \zeta_4
    - \frac{1048}{33}      \zeta_5
\right]\Break
+ \;     \alvp^{3}     \fl^{g}_{11}     n_f^{2}     { \frac{d^{abc}d^{abc}}{N_A} }\;
\left[
    - \frac{11661390042871}{128024064000} 
    - \frac{29154339}{107800}      \zeta_3
    + \frac{4408}{11}      \zeta_5
\right]\nonumber
{}
\end{eqnarray}
\begin{eqnarray}
\label{C2G12}
C^G_{2,12} & = & 
 \alvp n_f\;
\left[
    - \frac{8110049}{15135120} 
\right]\Break
+ \;     \alvp^{2}     C_A     n_f\;
\left[
    - \frac{39540735563386331}{10558130155372800} 
    - \frac{11}{13}      \zeta_3
\right]\Break
+ \;     \alvp^{2}     C_F     n_f\;
\left[
    - \frac{6258789011950819}{598533455520000} 
    + \frac{22}{13}      \zeta_3
\right]\Break
+ \;     \alvp^{3}     C_A     n_f^{2}\;
\left[
    + \frac{1102976289541525512630679}{62778008416037346432000} 
    - \frac{1574273}{8049132}      \zeta_3
\right]\Break
+ \;     \alvp^{3}     C_A^{2}     n_f\;
\left[
    - \frac{18295361799349570193991309242431}{102830377785469173455616000000} 
    + \frac{6726830671058291}{132943488678000}      \zeta_3
\right. \Break \left. \qquad \qquad
    + \frac{129763817}{16396380}      \zeta_4
    + \frac{20}{21}      \zeta_5
\right]\Break
+ \;     \alvp^{3}     C_F     C_A     n_f\;
\left[
    - \frac{9699486762665487150445457223019}{188522359273360151335296000000} 
    + \frac{789241683301607}{132943488678000}      \zeta_3
\right. \Break \left. \qquad \qquad
    - \frac{171207527}{16396380}      \zeta_4
    + \frac{4390}{273}      \zeta_5
\right]\Break
+ \;     \alvp^{3}     C_F     n_f^{2}\;
\left[
    + \frac{25016184592875203289560351501}{1346588280524001080966400000} 
    - \frac{13373036672}{7193911725}      \zeta_3
\right]\Break
+ \;     \alvp^{3}     C_F^{2}     n_f\;
\left[
    - \frac{194382513719914205037949320835327}{1555309464005221248516192000000} 
    + \frac{542538591728921}{33235872169500}      \zeta_3
\right. \Break \left. \qquad \qquad
    + \frac{2563}{1014}      \zeta_4
    - \frac{2220}{91}      \zeta_5
\right]\Break
+ \;     \alvp^{3}     \fl^{g}_{11}     n_f^{2}     { \frac{d^{abc}d^{abc}}{N_A} }\;
\left[
    - \frac{1656600471440498533}{10297935648000000} 
    - \frac{82799792129}{180589500}      \zeta_3
    + \frac{62436}{91}      \zeta_5
\right]\nonumber
{}
\end{eqnarray}
\begin{eqnarray}
\label{CLG10}
C^{G}_{L,10} & = & 
 \alvp n_f\;
\left[
    + \frac{2}{33} 
\right]\Break
+ \;     \alvp^{2}     C_A     n_f\;
\left[
    + \frac{2460678191}{1056547800} 
\right]\Break
+ \;     \alvp^{2}     C_F     n_f\;
\left[
    - \frac{509195549}{704365200} 
\right]\Break
+ \;     \alvp^{3}     C_A     n_f^{2}\;
\left[
    - \frac{140853814103239}{21965628762000} 
    - \frac{4}{33}      \zeta_3
\right]\Break
+ \;     \alvp^{3}     C_A^{2}     n_f\;
\left[
    + \frac{127219094296097749861}{1623699278087040000} 
    + \frac{5906419}{24012450}      \zeta_3
    - \frac{80}{11}      \zeta_5
\right]\Break
+ \;     \alvp^{3}     C_F     C_A     n_f\;
\left[
    - \frac{148061845621707638477}{2638511326891440000} 
    - \frac{210786157}{31216185}      \zeta_3
    + \frac{320}{11}      \zeta_5
\right]\Break
+ \;     \alvp^{3}     C_F     n_f^{2}\;
\left[
    + \frac{491491787586698683}{188465094777960000} 
    - \frac{136}{429}      \zeta_3
\right]\Break
+ \;     \alvp^{3}     C_F^{2}     n_f\;
\left[
    + \frac{9053411269935853949}{376930189555920000} 
    + \frac{860883476}{156080925}      \zeta_3
    - \frac{320}{11}      \zeta_5
\right]\Break
+ \;     \alvp^{3}     \fl^{g}_{11}     n_f^{2}     { \frac{d^{abc}d^{abc}}{N_A} }\;
\left[
    + \frac{112883693141257}{4526565120000} 
    + \frac{347273501}{4365900}      \zeta_3
    - \frac{1280}{11}      \zeta_5
\right]\nonumber
{}
\end{eqnarray}
\begin{eqnarray}
\label{CLG12}
C^G_{L,12} & = & 
 \alvp n_f\;
\left[
    + \frac{4}{91} 
\right] 
+ \;     \alvp^{2}     C_A     n_f\;
\left[
    + \frac{4978992299}{2685727044} 
\right] 
+ \;     \alvp^{2}     C_F     n_f\;
\left[
    - \frac{98593150597}{179847793125} 
\right]\Break
+ \;     \alvp^{3}     C_A     n_f^{2}\;
\left[
    - \frac{116919410865341069}{22683482755683750} 
    - \frac{8}{91}      \zeta_3
\right]\Break
+ \;     \alvp^{3}     C_A^{2}     n_f\;
\left[
    + \frac{96912603479263207273229}{1467873372990024000000} 
    - \frac{16493287}{717341625}      \zeta_3
    - \frac{480}{91}      \zeta_5
\right]\Break
+ \;     \alvp^{3}     C_F     C_A     n_f\;
\left[
    - \frac{2084160567995424969918773}{46709827690503978000000} 
    - \frac{9006563627}{2152024875}      \zeta_3
    + \frac{1920}{91}      \zeta_5
\right]\Break
+ \;     \alvp^{3}     C_F     n_f^{2}\;
\left[
    + \frac{49342300647723867029}{24328035255470821875} 
    - \frac{1696}{6825}      \zeta_3
\right]\Break
+ \;     \alvp^{3}     C_F^{2}     n_f\;
\left[
    + \frac{281711596115081884853551}{15569942563501326000000} 
    + \frac{2347741862}{717341625}      \zeta_3
    - \frac{1920}{91}      \zeta_5
\right]\Break
+ \;     \alvp^{3}     \fl^{g}_{11}     n_f^{2}     { \frac{d^{abc}d^{abc}}{N_A} }\;
\left[
    + \frac{2232852976776993919}{56638646064000000} 
    + \frac{115891012697}{993242250}      \zeta_3
    - \frac{15776}{91}      \zeta_5
\right]\nonumber
{}
\end{eqnarray}

\section{Results for the moments of $F_3$}

\begin{eqnarray}
\label{gamma3}
\gamma^{\mathrm{ns}}_{3} & = & 
 \alvp C_F\;
\left[
    + \frac{25}{6} 
\right] 
+ \;     \alvp^{2}     C_F     C_A\;
\left[
    + \frac{535}{27} 
\right] 
+ \;     \alvp^{2}     C_F     n_f\;
\left[
    - \frac{415}{108} 
\right] 
+ \;     \alvp^{2}     C_F^{2}\;
\left[
    - \frac{2035}{432} 
\right]\Break
+ \;     \alvp^{3}     C_F     C_A     n_f\;
\left[
    - \frac{62249}{3888} 
    - \frac{100}{3}      \zeta_3
\right] 
+ \;     \alvp^{3}     C_F     C_A^{2}\;
\left[
    + \frac{889433}{7776} 
    + \frac{55}{3}      \zeta_3
\right]\Break
+ \;     \alvp^{3}     C_F     n_f^{2}\;
\left[
    - \frac{2569}{1944} 
\right] 
+ \;     \alvp^{3}     C_F^{2}     C_A\;
\left[
    - \frac{311213}{15552} 
    - 55     \zeta_3
\right]\Break
+ \;     \alvp^{3}     C_F^{2}     n_f\;
\left[
    - \frac{203627}{7776} 
    + \frac{100}{3}      \zeta_3
\right] 
+ \;     \alvp^{3}     C_F^{3}\;
\left[
    - \frac{244505}{15552} 
    + \frac{110}{3}      \zeta_3
\right]\Break
+ \;     \alvp^{3}     n_f     { \frac{d^{abc}d^{abc}}{n} }\;
\left[
    + \frac{205}{288} 
\right]\nonumber
{}
\end{eqnarray}
\begin{eqnarray}
\label{gamma5}
\gamma^{\mathrm{ns}}_{5} & = & 
 \alvp C_F\;
\left[
    + \frac{91}{15} 
\right] 
+ \;     \alvp^{2}     C_F     C_A\;
\left[
    + \frac{73223}{2700} 
\right] 
+ \;     \alvp^{2}     C_F     n_f\;
\left[
    - \frac{7783}{1350} 
\right] 
+ \;     \alvp^{2}     C_F^{2}\;
\left[
    - \frac{2891}{500} 
\right]\Break
+ \;     \alvp^{3}     C_F     C_A     n_f\;
\left[
    - \frac{38587}{2000} 
    - \frac{728}{15}      \zeta_3
\right] 
+ \;     \alvp^{3}     C_F     C_A^{2}\;
\left[
    + \frac{305342801}{1944000} 
    + \frac{1414}{75}      \zeta_3
\right]\Break
+ \;     \alvp^{3}     C_F     n_f^{2}\;
\left[
    - \frac{215621}{121500} 
\right] 
+ \;     \alvp^{3}     C_F^{2}     C_A\;
\left[
    - \frac{63892213}{2430000} 
    - \frac{1414}{25}      \zeta_3
\right]\Break
+ \;     \alvp^{3}     C_F^{2}     n_f\;
\left[
    - \frac{5494973}{135000} 
    + \frac{728}{15}      \zeta_3
\right] 
+ \;     \alvp^{3}     C_F^{3}\;
\left[
    - \frac{51831073}{3037500} 
    + \frac{2828}{75}      \zeta_3
\right]\Break
+ \;     \alvp^{3}     n_f     { \frac{d^{abc}d^{abc}}{n} }\;
\left[
    + \frac{931}{4050} 
\right]\nonumber
{}
\end{eqnarray}
\begin{eqnarray}
\label{gamma7}
\gamma^{\mathrm{ns}}_{7} & = & 
 \alvp C_F\;
\left[
    + \frac{1027}{140} 
\right] 
+ \;     \alvp^{2}     C_F     C_A\;
\left[
    + \frac{67710257}{2116800} 
\right] 
+ \;     \alvp^{2}     C_F     n_f\;
\left[
    - \frac{3745727}{529200} 
\right]\Break
+ \;     \alvp^{2}     C_F^{2}\;
\left[
    - \frac{106801937}{16464000} 
\right] 
+ \;     \alvp^{3}     C_F     C_A     n_f\;
\left[
    - \frac{2257057261}{106686720} 
    - \frac{2054}{35}      \zeta_3
\right]\Break
+ \;     \alvp^{3}     C_F     C_A^{2}\;
\left[
    + \frac{219582793861}{1185408000} 
    + \frac{92741}{4900}      \zeta_3
\right] 
+ \;     \alvp^{3}     C_F     n_f^{2}\;
\left[
    - \frac{1369936511}{666792000} 
\right]\Break
+ \;     \alvp^{3}     C_F^{2}     C_A\;
\left[
    - \frac{629922436973}{20744640000} 
    - \frac{278223}{4900}      \zeta_3
\right] 
+ \;     \alvp^{3}     C_F^{2}     n_f\;
\left[
    - \frac{3150205788689}{62233920000} 
    + \frac{2054}{35}      \zeta_3
\right]\Break
+ \;     \alvp^{3}     C_F^{3}\;
\left[
    - \frac{151689902637457}{8712748800000} 
    + \frac{92741}{2450}      \zeta_3
\right] 
+ \;     \alvp^{3}     n_f     { \frac{d^{abc}d^{abc}}{n} }\;
\left[
    + \frac{56527729}{508032000} 
\right]\nonumber
{}
\end{eqnarray}
\begin{eqnarray}
\label{gamma9}
\gamma^{\mathrm{ns}}_{9} & = & 
 \alvp C_F\;
\left[
    + \frac{1045}{126} 
\right] 
+ \;     \alvp^{2}     C_F     C_A\;
\left[
    + \frac{242855129}{6804000} 
\right] 
+ \;     \alvp^{2}     C_F     n_f\;
\left[
    - \frac{19247947}{2381400} 
\right]\Break
+ \;     \alvp^{2}     C_F^{2}\;
\left[
    - \frac{6993510271}{1000188000} 
\right] 
+ \;     \alvp^{3}     C_F     C_A     n_f\;
\left[
    - \frac{63405201707}{2813028750} 
    - \frac{4180}{63}      \zeta_3
\right]\Break
+ \;     \alvp^{3}     C_F     C_A^{2}\;
\left[
    + \frac{744184331602493}{3600676800000} 
    + \frac{1253219}{66150}      \zeta_3
\right] 
+ \;     \alvp^{3}     C_F     n_f^{2}\;
\left[
    - \frac{20297329837}{9001692000} 
\right]\Break
+ \;     \alvp^{3}     C_F^{2}     C_A\;
\left[
    - \frac{5022136344247}{150028200000} 
    - \frac{1253219}{22050}      \zeta_3
\right] 
+ \;     \alvp^{3}     C_F^{2}     n_f\;
\left[
    - \frac{1630263834317}{28005264000} 
    + \frac{4180}{63}      \zeta_3
\right]\Break
+ \;     \alvp^{3}     C_F^{3}\;
\left[
    - \frac{13829238556849837}{793949234400000} 
    + \frac{1253219}{33075}      \zeta_3
\right] 
+ \;     \alvp^{3}     n_f     { \frac{d^{abc}d^{abc}}{n} }\;
\left[
    + \frac{13172190779}{200037600000} 
\right]\nonumber
{}
\end{eqnarray}
\begin{eqnarray}
\label{gamma11}
\gamma^{\mathrm{ns}}_{11} & = & 
 \alvp C_F\;
\left[
    + \frac{31408}{3465} 
\right] 
+ \;     \alvp^{2}     C_F     C_A\;
\left[
    + \frac{1448235599}{37422000} 
\right] 
+ \;     \alvp^{2}     C_F     n_f\;
\left[
    - \frac{512808781}{57629880} 
\right]\Break
+ \;     \alvp^{2}     C_F^{2}\;
\left[
    - \frac{2454220717793}{332812557000} 
\right] 
+ \;     \alvp^{3}     C_F     C_A     n_f\;
\left[
    - \frac{1031510572686647}{43568189280000} 
    - \frac{251264}{3465}      \zeta_3
\right]\Break
+ \;     \alvp^{3}     C_F     C_A^{2}\;
\left[
    + \frac{390549244457621303}{1742727571200000} 
    + \frac{151689577}{8004150}      \zeta_3
\right] 
+ \;     \alvp^{3}     C_F     n_f^{2}\;
\left[
    - \frac{28869611542843}{11981252052000} 
\right]\Break
+ \;     \alvp^{3}     C_F^{2}     C_A\;
\left[
    - \frac{43074459020106191}{1198125205200000} 
    - \frac{151689577}{2668050}      \zeta_3
\right]\Break
+ \;     \alvp^{3}     C_F^{2}     n_f\;
\left[
    - \frac{1188145134622636787}{18451128160080000} 
    + \frac{251264}{3465}      \zeta_3
\right]\Break
+ \;     \alvp^{3}     C_F^{3}\;
\left[
    - \frac{221430869576690083141}{12786631814935440000} 
    + \frac{151689577}{4002075}      \zeta_3
\right]\Break
+ \;     \alvp^{3}     n_f     { \frac{d^{abc}d^{abc}}{n} }\;
\left[
    + \frac{5083969985783}{116181838080000} 
\right]\nonumber
{}
\end{eqnarray}
\begin{eqnarray}
\label{gamma13}
\gamma^{\mathrm{ns}}_{13} & = & 
 \alvp C_F\;
\left[
    + \frac{874733}{90090} 
\right] 
+ \;     \alvp^{2}     C_F     C_A\;
\left[
    + \frac{31236494566127}{757512756000} 
\right] 
+ \;     \alvp^{2}     C_F     n_f\;
\left[
    - \frac{93360116539}{9739449720} 
\right]\Break
+ \;     \alvp^{2}     C_F^{2}\;
\left[
    - \frac{22445960639039759}{2924756750916000} 
\right]\Break
+ \;     \alvp^{3}     C_F     C_A     n_f\;
\left[
    - \frac{90849626920977361109}{3685193506154160000} 
    - \frac{3498932}{45045}      \zeta_3
\right]\Break
+ \;     \alvp^{3}     C_F     C_A^{2}\;
\left[
    + \frac{41070753377638233401027}{171975696953860800000} 
    + \frac{3662719609}{193243050}      \zeta_3
\right]\Break
+ \;     \alvp^{3}     C_F     n_f^{2}\;
\left[
    - \frac{66727681292862571}{26322810758244000} 
\right]\Break
+ \;     \alvp^{3}     C_F^{2}     C_A\;
\left[
    - \frac{1400874681707762602284653}{36888786996603141600000} 
    - \frac{3662719609}{64414350}      \zeta_3
\right]\Break
+ \;     \alvp^{3}     C_F^{2}     n_f\;
\left[
    - \frac{36688336888519925613757}{526982671380044880000} 
    + \frac{3498932}{45045}      \zeta_3
\right]\Break
+ \;     \alvp^{3}     C_F^{3}\;
\left[
    - \frac{40795447722180713788820819}{2373793443231412161960000} 
    + \frac{3662719609}{96621525}      \zeta_3
\right]\Break
+ \;     \alvp^{3}     n_f     { \frac{d^{abc}d^{abc}}{n} }\;
\left[
    + \frac{62160363128061559}{1984334964852240000} 
\right]\nonumber
{}
\end{eqnarray}

\begin{eqnarray}
\label{C1}
C^{\mathrm{ns}}_1 & = & 
1\;
{}
+ \; \alvp C_F\;
\left[
    - 3
\right]\Break
+ \;     \alvp^{2}     C_F     C_A\;
\left[
    - 23
\right] 
+ \;     \alvp^{2}     C_F     n_f\;
\left[
    + 4
\right] 
+ \;     \alvp^{2}     C_F^{2}\;
\left[
    + \frac{21}{2} 
\right]\Break
+ \;     \alvp^{3}     C_F     C_A     n_f\;
\left[
    + \frac{3535}{27} 
    + 24     \zeta_3
    - \frac{80}{3}      \zeta_5
\right] 
+ \;     \alvp^{3}     C_F     C_A^{2}\;
\left[
    - \frac{10874}{27} 
    + \frac{440}{3}      \zeta_5
\right]\Break
+ \;     \alvp^{3}     C_F     n_f^{2}\;
\left[
    - \frac{230}{27} 
\right] 
+ \;     \alvp^{3}     C_F^{2}     C_A\;
\left[
    + \frac{1241}{9} 
    - \frac{176}{3}      \zeta_3
\right] 
+ \;     \alvp^{3}     C_F^{2}     n_f\;
\left[
    - \frac{133}{18} 
    - \frac{40}{3}      \zeta_3
\right]\Break
+ \;     \alvp^{3}     C_F^{3}\;
\left[
    - \frac{3}{2} 
\right] 
+ \;     \alvp^{3}     n_f     { \frac{d^{abc}d^{abc}}{n} }\;
\left[
    - \frac{11}{3} 
    + 8     \zeta_3
\right]\nonumber
{}
\end{eqnarray}
\begin{eqnarray}
\label{C3}
C^{\mathrm{ns}}_{3} & = & 
1\;
{}
+ \; \alvp C_F\;
\left[
    + \frac{5}{4} 
\right]\Break
+ \;     \alvp^{2}     C_F     C_A\;
\left[
    + \frac{5209}{144} 
    - 33     \zeta_3
\right] 
+ \;     \alvp^{2}     C_F     n_f\;
\left[
    - \frac{4369}{864} 
\right] 
+ \;     \alvp^{2}     C_F^{2}\;
\left[
    - \frac{34763}{10368} 
    + 16     \zeta_3
\right]\Break
+ \;     \alvp^{3}     C_F     C_A     n_f\;
\left[
    - \frac{24877649}{699840} 
    + \frac{18539}{405}      \zeta_3
    - \frac{50}{3}      \zeta_4
\right]\Break
+ \;     \alvp^{3}     C_F     C_A^{2}\;
\left[
    + \frac{92517547}{699840} 
    - \frac{225337}{405}      \zeta_3
    + \frac{55}{6}      \zeta_4
    + 390     \zeta_5
\right]\Break
+ \;     \alvp^{3}     C_F     n_f^{2}\;
\left[
    - \frac{12125}{69984} 
    + \frac{100}{81}      \zeta_3
\right] 
+ \;     \alvp^{3}     C_F^{2}     C_A\;
\left[
    + \frac{222157399}{559872} 
    + \frac{206}{9}      \zeta_3
    - \frac{55}{2}      \zeta_4
    - 260     \zeta_5
\right]\Break
+ \;     \alvp^{3}     C_F^{2}     n_f\;
\left[
    - \frac{4292227}{34992} 
    + \frac{527}{9}      \zeta_3
    + \frac{50}{3}      \zeta_4
\right]\Break
+ \;     \alvp^{3}     C_F^{3}\;
\left[
    - \frac{1041473}{373248} 
    + \frac{2183}{324}      \zeta_3
    + \frac{55}{3}      \zeta_4
    - 40     \zeta_5
\right] 
+ \;     \alvp^{3}     n_f     { \frac{d^{abc}d^{abc}}{n} }\;
\left[
    - \frac{19477}{5184} 
    + 5     \zeta_3
\right]\nonumber
{}
\end{eqnarray}
\begin{eqnarray}
\label{C5}
C^{\mathrm{ns}}_{5} & = & 
1\;
{}
+ \; \alvp C_F\;
\left[
    + \frac{523}{90} 
\right]\Break
+ \;     \alvp^{2}     C_F     C_A\;
\left[
    + \frac{30312449}{324000} 
    - \frac{276}{5}      \zeta_3
\right] 
+ \;     \alvp^{2}     C_F     n_f\;
\left[
    - \frac{2356859}{162000} 
\right]\Break
+ \;     \alvp^{2}     C_F^{2}\;
\left[
    - \frac{14729261}{1620000} 
    + \frac{188}{5}      \zeta_3
\right]\Break
+ \;     \alvp^{3}     C_F     C_A     n_f\;
\left[
    - \frac{33317596477}{122472000} 
    + \frac{1413442}{14175}      \zeta_3
    - \frac{364}{15}      \zeta_4
\right]\Break
+ \;     \alvp^{3}     C_F     C_A^{2}\;
\left[
    + \frac{142639414763}{163296000} 
    - \frac{30424087}{28350}      \zeta_3
    + \frac{707}{75}      \zeta_4
    + 548     \zeta_5
\right]\Break
+ \;     \alvp^{3}     C_F     n_f^{2}\;
\left[
    + \frac{784091}{54000} 
    + \frac{728}{405}      \zeta_3
\right]\Break
+ \;     \alvp^{3}     C_F^{2}     C_A\;
\left[
    + \frac{179060877287}{255150000} 
    - \frac{512039}{7875}      \zeta_3
    - \frac{707}{25}      \zeta_4
    - 180     \zeta_5
\right]\Break
+ \;     \alvp^{3}     C_F^{2}     n_f\;
\left[
    - \frac{28068155797}{127575000} 
    + \frac{205712}{4725}      \zeta_3
    + \frac{364}{15}      \zeta_4
\right]\Break
+ \;     \alvp^{3}     C_F^{3}\;
\left[
    - \frac{29707273013}{2187000000} 
    + \frac{4045486}{10125}      \zeta_3
    + \frac{1414}{75}      \zeta_4
    - 376     \zeta_5
\right]\Break
+ \;     \alvp^{3}     n_f     { \frac{d^{abc}d^{abc}}{n} }\;
\left[
    - \frac{24801551}{17496000} 
    + \frac{86}{45}      \zeta_3
\right]\nonumber
{}
\end{eqnarray}
\begin{eqnarray}
\label{C7}
C^{\mathrm{ns}}_{7} & = & 
1\;
{}
+ \; \alvp C_F\;
\left[
    + \frac{48091}{5040} 
\right]\Break
+ \;     \alvp^{2}     C_F     C_A\;
\left[
    + \frac{502042084559}{3556224000} 
    - \frac{4917}{70}      \zeta_3
\right] 
+ \;     \alvp^{2}     C_F     n_f\;
\left[
    - \frac{20352710029}{889056000} 
\right]\Break
+ \;     \alvp^{2}     C_F^{2}\;
\left[
    + \frac{972223395949}{248935680000} 
    + \frac{1836}{35}      \zeta_3
\right]\Break
+ \;     \alvp^{3}     C_F     C_A     n_f\;
\left[
    - \frac{3373161364214023}{6721263360000} 
    + \frac{4838581}{33075}      \zeta_3
    - \frac{1027}{35}      \zeta_4
\right]\Break
+ \;     \alvp^{3}     C_F     C_A^{2}\;
\left[
    + \frac{5386502616041647}{3360631680000} 
    - \frac{134407981}{88200}      \zeta_3
    + \frac{92741}{9800}      \zeta_4
    + \frac{14108}{21}      \zeta_5
\right]\Break
+ \;     \alvp^{3}     C_F     n_f^{2}\;
\left[
    + \frac{9986153999891}{336063168000} 
    + \frac{2054}{945}      \zeta_3
\right]\Break
+ \;     \alvp^{3}     C_F^{2}     C_A\;
\left[
    + \frac{2316127855865945089}{1881953740800000} 
    - \frac{450138923}{4630500}      \zeta_3
    - \frac{278223}{9800}      \zeta_4
    - \frac{4118}{21}      \zeta_5
\right]\Break
+ \;     \alvp^{3}     C_F^{2}     n_f\;
\left[
    - \frac{3455733053370931}{9801842400000} 
    + \frac{483863}{26460}      \zeta_3
    + \frac{1027}{35}      \zeta_4
\right]\Break
+ \;     \alvp^{3}     C_F^{3}\;
\left[
    - \frac{906919428644261623}{14637417984000000} 
    + \frac{27358836137}{37044000}      \zeta_3
    + \frac{92741}{4900}      \zeta_4
    - \frac{11224}{21}      \zeta_5
\right]\Break
+ \;     \alvp^{3}     n_f     { \frac{d^{abc}d^{abc}}{n} }\;
\left[
    - \frac{1521158314519}{1920360960000} 
    + \frac{529}{504}      \zeta_3
\right]\nonumber
{}
\end{eqnarray}
\begin{eqnarray}
\label{C9}
C^{\mathrm{ns}}_{9} & = & 
1\;
{}
+ \; \alvp C_F\;
\left[
    + \frac{17761}{1400} 
\right]\Break
+ \;     \alvp^{2}     C_F     C_A\;
\left[
    + \frac{6247950073621}{34292160000} 
    - \frac{2858}{35}      \zeta_3
\right] 
+ \;     \alvp^{2}     C_F     n_f\;
\left[
    - \frac{363260060687}{12002256000} 
\right]\Break
+ \;     \alvp^{2}     C_F^{2}\;
\left[
    + \frac{26950029280781}{1008189504000} 
    + \frac{6698}{105}      \zeta_3
\right]\Break
+ \;     \alvp^{3}     C_F     C_A     n_f\;
\left[
    - \frac{1532454569897512927}{2138802019200000} 
    + \frac{121711969}{654885}      \zeta_3
    - \frac{2090}{63}      \zeta_4
\right]\Break
+ \;     \alvp^{3}     C_F     C_A^{2}\;
\left[
    + \frac{15496996833630287207}{6805279152000000} 
    - \frac{7620848351}{3969000}      \zeta_3
    + \frac{1253219}{132300}      \zeta_4
    + \frac{49634}{63}      \zeta_5
\right]\Break
+ \;     \alvp^{3}     C_F     n_f^{2}\;
\left[
    + \frac{3023214594598871}{68052791520000} 
    + \frac{4180}{1701}      \zeta_3
\right]\Break
+ \;     \alvp^{3}     C_F^{2}     C_A\;
\left[
    + \frac{1111314791170240118909}{571643448768000000} 
    - \frac{4449188411}{41674500}      \zeta_3
    - \frac{1253219}{44100}      \zeta_4
    - \frac{2678}{9}      \zeta_5
\right]\Break
+ \;     \alvp^{3}     C_F^{2}     n_f\;
\left[
    - \frac{80521228766527611043}{157201948411200000} 
    - \frac{22156972}{3274425}      \zeta_3
    + \frac{2090}{63}      \zeta_4
\right]\Break
+ \;     \alvp^{3}     C_F^{3}\;
\left[
    - \frac{388781612300063841847}{4001504141376000000} 
    + \frac{95897878894}{93767625}      \zeta_3
    + \frac{1253219}{66150}      \zeta_4
    - \frac{5092}{9}      \zeta_5
\right]\Break
+ \;     \alvp^{3}     n_f     { \frac{d^{abc}d^{abc}}{n} }\;
\left[
    - \frac{390642677252603}{756142128000000} 
    + \frac{95839}{141750}      \zeta_3
\right]\nonumber
{}
\end{eqnarray}
\begin{eqnarray}
\label{C11}
C^{\mathrm{ns}}_{11} & = & 
1\;
{}
+ \; \alvp C_F\;
\left[
    + \frac{12815983}{831600} 
\right]\Break
+ \;     \alvp^{2}     C_F     C_A\;
\left[
    + \frac{56653343996617}{259334460000} 
    - \frac{104947}{1155}      \zeta_3
\right] 
+ \;     \alvp^{2}     C_F     n_f\;
\left[
    - \frac{117825956269669}{3195000547200} 
\right]\Break
+ \;     \alvp^{2}     C_F^{2}\;
\left[
    + \frac{25436786950269217}{461278204002000} 
    + \frac{84262}{1155}      \zeta_3
\right]\Break
+ \;     \alvp^{3}     C_F     C_A     n_f\;
\left[
    - \frac{5339200983355026311}{5825689309440000} 
    + \frac{206111663893}{936485550}      \zeta_3
    - \frac{125632}{3465}      \zeta_4
\right]\Break
+ \;     \alvp^{3}     C_F     C_A^{2}\;
\left[
    + \frac{5458605391363920639632689}{1884027922672896000000} 
    - \frac{8531884151173}{3745942200}      \zeta_3
    + \frac{151689577}{16008300}      \zeta_4
\right. \Break \left. \qquad \qquad
    + \frac{209072}{231}      \zeta_5
\right]\Break
+ \;     \alvp^{3}     C_F     n_f^{2}\;
\left[
    + \frac{58243229827221303967}{996360920644320000} 
    + \frac{251264}{93555}      \zeta_3
\right]\Break
+ \;     \alvp^{3}     C_F^{2}     C_A\;
\left[
    + \frac{37894668775960020571538737}{13600326566794968000000} 
    - \frac{52724308943557}{576875098800}      \zeta_3
    - \frac{151689577}{5336100}      \zeta_4
\right. \Break \left. \qquad \qquad
    - \frac{46958}{99}      \zeta_5
\right]\Break
+ \;     \alvp^{3}     C_F^{2}     n_f\;
\left[
    - \frac{7504061852485890170936687}{10880261253435974400000} 
    - \frac{2560572011}{85135050}      \zeta_3
    + \frac{125632}{3465}      \zeta_4
\right]\Break
+ \;     \alvp^{3}     C_F^{3}\;
\left[
    - \frac{358252596866786733349510147}{3544454339100103968000000} 
    + \frac{156213323360401}{124804708875}      \zeta_3
    + \frac{151689577}{8004150}      \zeta_4
\right. \Break \left. \qquad \qquad
    - \frac{49124}{99}      \zeta_5
\right]\Break
+ \;     \alvp^{3}     n_f     { \frac{d^{abc}d^{abc}}{n} }\;
\left[
    - \frac{1182009270543683429}{3220560551577600000} 
    + \frac{6758}{14175}      \zeta_3
\right]\nonumber
{}
\end{eqnarray}
\begin{eqnarray}
\label{C13}
C^{\mathrm{ns}}_{13} & = & 
1\;
{}
+ \; \alvp C_F\;
\left[
    + \frac{84292133}{4729725} 
\right]\Break
+ \;     \alvp^{2}     C_F     C_A\;
\left[
    + \frac{137082775015804598489}{545954593504320000} 
    - \frac{1480186}{15015}      \zeta_3
\right]\Break
+ \;     \alvp^{2}     C_F     n_f\;
\left[
    - \frac{301032882286150933}{7019416202198400} 
\right]\Break
+ \;     \alvp^{2}     C_F^{2}\;
\left[
    + \frac{5233449307353261454457}{60226591014862272000} 
    + \frac{1210906}{15015}      \zeta_3
\right]\Break
+ \;     \alvp^{3}     C_F     C_A     n_f\;
\left[
    - \frac{879328426004243254681619899}{796797799126627858560000} 
    + \frac{3046809425749}{12174312150}      \zeta_3
    - \frac{1749466}{45045}      \zeta_4
\right]\Break
+ \;     \alvp^{3}     C_F     C_A^{2}\;
\left[
    + \frac{645026845788679520269283647601}{185919486462879833664000000} 
    - \frac{37039408217872}{14203364175}      \zeta_3
\right. \Break \left. \qquad \qquad
    + \frac{3662719609}{386486100}      \zeta_4
    + \frac{3086186}{3003}      \zeta_5
\right]\Break
+ \;     \alvp^{3}     C_F     n_f^{2}\;
\left[
    + \frac{2043359170316436833750887}{28457064254522423520000} 
    + \frac{3498932}{1216215}      \zeta_3
\right]\Break
+ \;     \alvp^{3}     C_F^{2}     C_A\;
\left[
    + \frac{9081431864860519767694223551709}{2441616113038023938016000000} 
    - \frac{2292102879924479}{48745945848600}      \zeta_3
\right. \Break \left. \qquad \qquad
    - \frac{3662719609}{128828700}      \zeta_4
    - \frac{6487178}{9009}      \zeta_5
\right]\Break
+ \;     \alvp^{3}     C_F^{2}     n_f\;
\left[
    - \frac{19506028401623230293412698913}{22196510118527490345600000} 
    - \frac{313559848919}{6087156075}      \zeta_3
    + \frac{1749466}{45045}      \zeta_4
\right]\Break
+ \;     \alvp^{3}     C_F^{3}\;
\left[
    - \frac{298680730292949187763497559489677}{4277101026014358433419528000000} 
    + \frac{142217314466274683}{99707616508500}      \zeta_3
\right. \Break \left. \qquad \qquad
    + \frac{3662719609}{193243050}      \zeta_4
    - \frac{235868}{693}      \zeta_5
\right]\Break
+ \;     \alvp^{3}     n_f     { \frac{d^{abc}d^{abc}}{n} }\;
\left[
    - \frac{65696095155620706347863}{238358315978051068800000} 
    + \frac{11802932}{33108075}      \zeta_3
\right]\nonumber
{}
\end{eqnarray}
\end{appendix}

\end{document}